\documentclass[twocolumn]{aastex631}
\usepackage{amsmath}

\newcommand{\kms}{\rm km~s^{-1}}
\newcommand{\kmsmpc}{\rm km~s^{-1}~Mpc^{-1}}
\newcommand{\dn}{D_{n}4000}

\usepackage{color}
\usepackage{multirow}
\usepackage{amsmath}
\usepackage{appendix}
\usepackage{hhline}
\usepackage{hyperref} 

\begin{document}

\title{Quiescent Galaxy Size, Velocity Dispersion, and Dynamical Mass Evolution}

\author{Ivana Damjanov}
\affil{Department of Astronomy and Physics, Saint Mary's University, 923 Robie Street, Halifax, NS B3H 3C3, Canada; \href{mailto:Ivana.Damjanov@smu.ca}{Ivana.Damjanov@smu.ca}}
\affil{Canada Research Chair in Astronomy and Astrophysics, Tier II}
\author{Jubee Sohn}
\affil{Harvard-Smithsonian Center for Astrophysics, 60 Garden Street, Cambridge, MA 02138, USA}
\author{Yousuke Utsumi}
\affil{SLAC National Accelerator Laboratory, Menlo Park, CA 94025, USA}
\affil{Kavli Institute for Particle Astrophysics and Cosmology, Stanford University, Stanford, CA 94305, USA}
\author{Margaret J. Geller}
\affil{Harvard-Smithsonian Center for Astrophysics, 60 Garden Street, Cambridge, MA 02138, USA}
\author{Ian Dell'Antonio}
\affil{Department of Physics, Brown University, Box 1843, Providence, RI 02912, USA}

\begin{abstract}

We use surveys covering the redshift range $0.05 < z < 3.8$
to explore quiescent galaxy scaling relations and the redshift evolution of
the velocity dispersion, size, and dynamical mass at fixed  stellar mass. For redshift $z < 0.6$ we derive mass limited samples and demonstrate that these large samples enhance constraints on the evolution of the quiescent population. The constraints include 2985 new velocity dispersions from the SHELS F2 survey
\citep{Geller14}. In contrast with the known substantial evolution of size with redshift, evolution in the velocity dispersion is negligible. The dynamical-to-stellar mass ratio increases significantly as the universe ages in agreement with recent results that combine high redshift data with the SDSS.
Like other investigators, we interpret this result as an indication that the dark matter fraction within the effective radius increases as a result of the impact of the minor mergers that are responsible for size growth. We emphasize that dense redshift surveys covering the range $0.07 < z < 1$ along with strong and weak lensing measurements could remove many ambiguities in evolutionary studies of the quiescent population.

\end{abstract}
 
\section{Introduction}

Understanding the evolution of the deceptively simple quiescent galaxy population presents an array of observational and theoretical challenges. Measured sizes, velocity dispersions and dynamical masses as a function of stellar mass provide potential constraints and insights. We review the projections of the stellar mass fundamental plane and the relationship between dynamical and stellar mass. We explore the evolution of these relations constrained by the inclusion of a dense complete sample at redshift $0.2 < z < 0.6$.

The size-stellar mass relation is a projection of the classic stellar mass fundamental plane \citep[e.g.,][]{Hyde2009a,Bezanson2013,Zahid2016a}. For galaxies with stellar mass $M_\ast\gtrsim10^{10}\, M_\sun$, the exponent of the $R_{e} = A\times M_\ast^{\alpha}$ relation is typically $0.6\lesssim\alpha\lesssim0.7$ with only a marginal dependence on redshift. On the other hand, the normalization $A$ of this relation evolves substantially and this size evolution of the quiescent population is well documented \citep[][and others]{Trujillo2006,Trujillo2007,Toft2007,vanDokkum2010,Williams2010,Damjanov2011,Ryan2012,Newman12,Cassata2013,Huertas-Company2013,vdW14,Faisst2017,Damjanov19,Mowla2019,Mosleh2020,Yang2021,kawinwanichakij_hyper_2021}. Quiescent galaxies of fixed stellar mass grow by a factor $2.5-4$ from redshift of $z\sim 1.5$ to the present. Minor merger driven growth \citep[e.g.,][]{Naab2009} is increasingly widely accepted as the dominant driver of this evolution.

The relation between central velocity dispersion $\sigma_e$ and stellar mass, also a projection of the stellar mass fundamental plane, yields further insights into the quiescent galaxy population and its evolution \citep{Gallazzi2006,Hyde09,Shankar2010,Aguerri2012, Belli2014, Zahid16, Napolitano2020}. \citet{Zahid16} show that $\sigma_e \propto M_{*}^{1/3}$; the slope is essentially invariant for $z \lesssim 0.7$. Remarkably, the  slope is identical to the slope of the analogous scaling in simulations of dark matter halos \citep{Evrard08, Posti14}. Evolution of the velocity dispersion of the  quiescent population at fixed stellar mass  is much less dramatic than the corresponding size evolution. Results in the literature range from essentially no evolution  to a small decrease of velocity dispersion as the universe ages for galaxies at fixed stellar mass \citep[e.g.,][]{Cenarro2009, vdS13, Cannarozzo20, Stockmann2020}. 

 Dynamical masses combine the velocity dispersion and size, possibly opening a window  on the dark matter fraction and quiescent galaxy evolution. Evaluation of the scaling between dynamical mass and stellar mass is a subtle task partly because of the observational selection effects \citep{Zahid17}. Variations in the surface brightness profiles (structural non-homology) of quiescent galaxies are an additional issue \citep{Bertin02,Cappellari06,Novak2012}. When these issues are taken into account, there is a nearly direct proportionality between stellar and dynamical mass \citep{Taylor10, Zahid17}.  The \citet{Bolton08} study of a sample of 53 strong lenses with redshift in the range $0.06<z<0.36$ demonstrates the correspondence between the lensing mass and dynamical mass. These lensing results suggest that the dynamical mass measures the total mass including the dark matter contribution. 
 
 Early studies at high redshift suggest negligible evolution in the dynamical-to-stellar mass ratio \citep{vdS13,Belli2014}. More recent observations of high redshift samples \citep[e.g.,][]{Stockmann2020, Mendel2020, Esdaile2021} imply that the ratio between the dynamical mass and the stellar mass increases as the universe ages. Systematic uncertainties in the computation of stellar masses complicate the interpretation of these results, but the consensus is that the dark matter fraction contained within the effective radius increases as the universe ages \citep{Hilz2013,Frigo17,Remus2017,Lovell2018}. Preferential deposition of material in the outer regions of galaxies by minor mergers potentially accounts for this evolution.

We use  dense SHELS F2 data \citep{Geller14} covering the redshift range $0.2 < z < 0.6$ combined with low redshift data from the Sloan Digital Sky Survey \citep[SDSS,][]{Abazajian2009,Ahumada20} to explore the relations among central stellar velocity dispersion, dynamical mass, and stellar mass at redshifts $\lesssim 0.6$. The SHELS survey of the F2 field of the Deep Lens Survey \citep{Wittman06} is complete to $r = 20.75$ and includes 10,848 individual galaxies with a redshift. Among quiescent galaxies in this sample, 3465/3861 have a central stellar velocity dispersion. \citet{Damjanov19} use the HSC imaging to derive sizes for galaxies in the SHELS survey, but they do not extend their analysis to the determination of dynamical masses. Here we use the HSC sizes to derive the dynamical masses. We use the available high redshift samples of \citet{vdS13}, \citet{Belli17}, \citet{Stockmann2020}, \citet{Mendel2020}, and \citet{Esdaile2021} to extend the analysis to explore the redshift range 0.8 - 3.8.

We demonstrate that dense complete samples of quiescent galaxies spanning the redshift range  $z\lesssim 0.6$ significantly enhance constraints on the evolution of size, velocity dispersion and dynamical mass as a function of stellar mass. We highlight both the agreement and some subtle tensions between these $z\lesssim 0.6$  results and the ones based on existing higher redshift samples. 

We review the SDSS data and SHELS F2 data in Section \ref{sec:DATA}. We also explore additional available high redshift samples in Section \ref{allsamples}. We revisit the scaling relations and evolution for quiescent galaxy velocity dispersion and size in Section \ref{scaleevolution}. Section \ref{dynmass} outlines the evolution of the relation between dynamical mass and stellar mass. We discuss the dark matter content (Section \ref{interpretation} and the potential impact of strong and weak lensing observations (Section \ref{lensing}). We conclude in Section \ref{conclusion}. We use $H_{0} = 70~\kmsmpc$, $\Omega_{m} = 0.3$, and $\Omega_{\Lambda} = 0.7$ throughout the paper. 

\section{The Data}\label{sec:DATA}

We use large samples from the SDSS and SHELS redshift surveys to study the scaling between dynamical and stellar mass from redshift 0.1 to 0.6. SDSS typifies the properties of galaxies at $z \lesssim 0.1$. The SHELS survey explores galaxy properties at a median $z \sim 0.3$. For a cleaner comparison between the SDSS shallow sample and SHELS, we reconstruct the catalog of galaxy redshifts in the SHELS F2 field based on SDSS DR16 photometry \citep{Ahumada20}. We follow \citet{vdS13} and \citet{vdS15} in sampling the SDSS as a low redshift baseline. For constraints at higher redshift we include samples from \citet{vdS13}, \citet{Belli17} \citet{Stockmann2020}, \citet{Mendel2020}, and \citet{Esdaile2021} (see Section \ref{allsamples}).

\subsection{Sloan Digital Sky Survey}\label{sec:SDSS}

Following \citet{vdS13} and \citet{vdS15}, we select 93,485 galaxies with $0.05 < z < 0.07$ from SDSS DR7 \citep{SDSSDR7}. As in our previous studies of the quiescent population we use the spectral indicator $\dn$ to define quiescence \citep{Zahid16,Zahid2016a,Zahid17,Sohn17,Damjanov18,Damjanov19}. The SDSS MPA/JHU catalog provides the $\dn$ data we use to segregate the population. There are 48,073 SDSS galaxies with $\dn > 1.6$ (we discuss the choice of a limiting $\dn$ in section \ref{D4000} below).

\begin{deluxetable*}{lcc}[!ht]
\tablecaption{Physical Property Sources for the SDSS and SHELS F2 Sample}
\label{Table1}
\tablecolumns{2}
\tabletypesize{\scriptsize}
\tablewidth{0pt}
\tablehead{\colhead{Physical Property } & \colhead{SDSS} & \colhead{F2}}
\startdata
Redshift                                  & SDSS/MPA                    & SDSS + MMT/Hectospec   \\
Stellar mass ($M_{*}$)                    & Le Phare (Petrosian)        & Le Phare (Petrosian) \\
Circularized effective radius ($R_{e}$)       & \citet{Simard11} (GIM2D)    & \citet{Damjanov19} (Source Extractor)      \\
S\'{e}rsic Index ($n$)                        & \citet{Simard11} (GIM2D)    & \citet{Damjanov19} (Source Extractor)       \\
K-correction                              & \citet{Blanton03}           & \citet{Blanton03}      \\
Velocity dispersion                       & Portsmouth-pPXF                 & ULySS                  \\
$\beta$                                   & \citet{Cappellari06}        & \citet{Cappellari06}   \\
Quiescent Galaxy ID                       & $\dn > 1.6$                 & $\dn > 1.6$            \\
\enddata
\end{deluxetable*}

\begin{figure*}[ht!]
\centering
\includegraphics[width=0.9\textwidth]{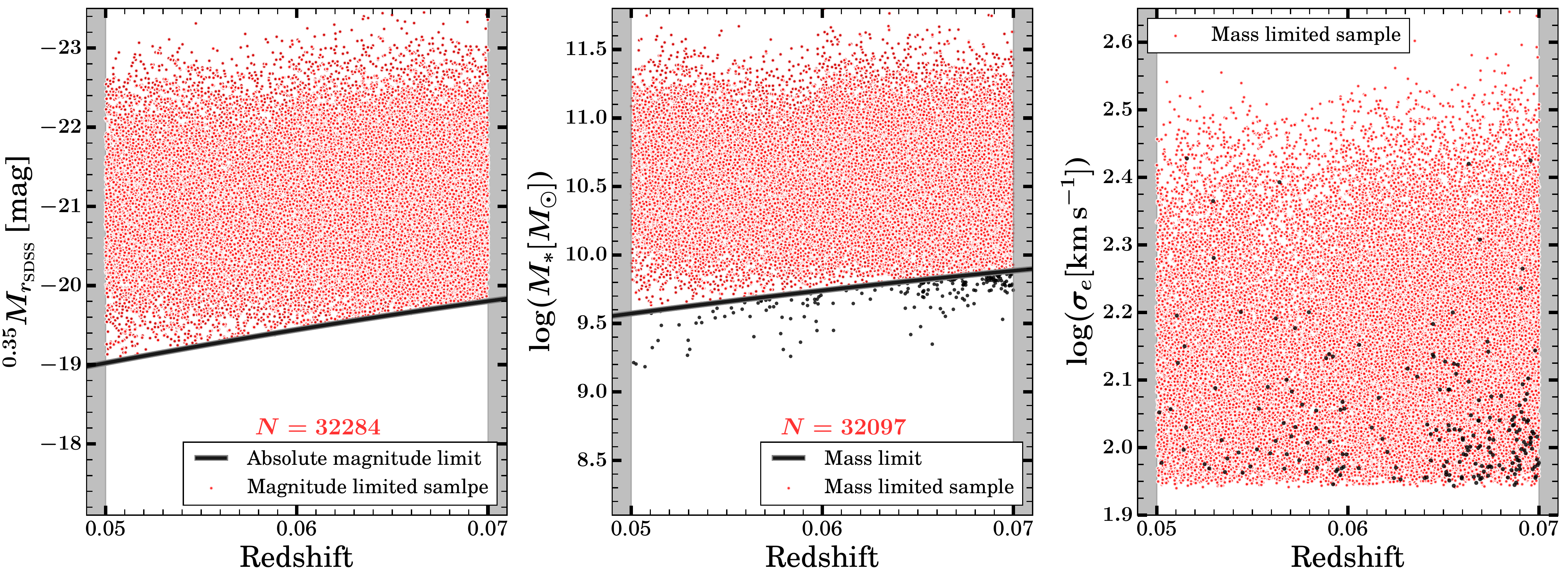}
\caption{Absolute $r-$band magnitude ({\it left} panel), stellar mass ({\it central} panel), and velocity dispersion ({\it right} panel) versus redshift for the $0.05 \leq z \leq 0.07$ quiescent SDSS Legacy Survey galaxy sample. The absolute magnitude limit (black solid line in the left panel), corresponding to the observed $r-$band magnitude limit of $r = 17.78$, translates to the stellar mass limit (black solid line in the central panel) by incorporating the average mass-to-light ratio for SDSS galaxies at a given redshift. Red points in the left panel indicate the subset of galaxies above the absolute magnitude completeness limit. Red points in the central panel and right indicate the subset of galaxies above the stellar mass completeness limit. Black points correspond to the galaxies that are above the absolute magnitude limit, but are not in the mass limited sample. The legends give the number of SDSS galaxies in the absolute magnitude and stellar mass limited samples. }
\label{fig1}
\end{figure*}

Figure \ref{fig1} shows characteristics of the SDSS quiescent sample. The left panel shows the absolute K-corrected $r-$band magnitude as a function of redshift. The black solid line indicates the absolute magnitude limit corresponding to the SDSS Main Galaxy Sample r-band completeness limit $r = 17.78$ \citep{Strauss02}. We shift the K-corrected absolute magnitudes to a redshift $z = 0.3$ for comparison with the deeper SHELS survey.

The middle panel shows the stellar mass as a function of redshift. In constructing Figure \ref{fig1}, we take stellar masses from the MPA/JHU catalog (Table \ref{Table1}). Table \ref{Table1} lists the physical parameters we compile from the SDSS and related value added catalogs. The black line shows the stellar mass completeness limit based on the typical mass-to-light ratio for quiescent galaxies at the SDSS limiting $r = 17.78$ at each redshift. A negligible fraction of galaxies lie above the absolute magnitude limit but fall below the stellar mass completeness threshold due to the spread in mass-to-light ratio at a given redshift (black points in the central panel). The sample is complete throughout the redshift range for $M_{*} > 10^{10}$ $M_{\odot}$.

Finally, the right hand panel shows the central velocity dispersion as a function of redshift. At a fixed absolute magnitude (or stellar mass) the spread in velocity dispersion is large. Thus neither magnitude limited nor stellar mass limited samples are velocity dispersion limited. \citet{Sohn17} discuss this issue in detail. The small number of galaxies from the absolute magnitude limited sample that are not in the mass limited sample have uniformly distributed velocity dispersion values (black points in the right panel). The Figure~\ref{fig1} legends in the left and central panels give the number of galaxies in the magnitude and mass limited samples, respectively.

\subsection{SHELS F2 Survey}

The Deep Lens Survey \citep{Wittman06} was a pioneering imaging project designed to detect weak lensing in 5 disjoint fields, each covering about four square degrees of the sky. In 2004 - 2009, \citet{Geller14} used the Hectospec 300-fiber instrument \citep{Fabricant05} mounted on the MMT to acquire spectroscopy for galaxies in the DLS F2 field centered at R.A. = 139.89$^\circ$ and Decl. = 30.00$^\circ$. \citet{Geller14} describe the observational approach to both photometry and spectroscopy in Section 2 of their paper. 

\begin{deluxetable*}{lcccccc}
\tablecaption{SHELS F2 Sample\tablenotemark{a}}
\label{Table2}
\tablecolumns{2}
\tabletypesize{\small}
\tablewidth{0pt}
\tablehead{\colhead{SDSS ID} & \colhead{R.A.} & \colhead{Dec} & \colhead{$r_\mathrm{petro}$} & \colhead{$z$} & \colhead{D$_n4000$} & \colhead{$\log(M_*/M_\sun)$}}
\startdata
1237664093976986160 & 138.70035 & 30.73643 & 19.213$\pm$0.037 & 0.27381$\pm$0.00015 & 1.623$\pm$0.065 & 10.772$_{-0.122}^{+0.096}$ \\
1237664093976986057 & 138.70127 & 30.65197 & 20.286$\pm$0.067 & 0.39838$\pm$0.00011 & 1.731$\pm$0.063 & 10.835$_{-0.171}^{+0.104}$ \\
1237664668965601390 & 138.70158 & 30.44852 & 18.506$\pm$0.019 & 0.12403$\pm$0.00010 & 1.263$\pm$0.024 & 10.247$_{-0.098}^{+0.134}$ \\
1237664669502538244 & 138.70187 & 30.81657 & 20.637$\pm$0.100 & 0.39797$\pm$0.00016 & 1.985$\pm$0.116 & 10.810$_{-0.142}^{+0.112}$ \\
1237664668965601456 & 138.70280 & 30.54220 & 18.820$\pm$0.045 & 0.11266$\pm$0.00009 & 1.926$\pm$0.141 & 10.040$_{-0.149}^{+0.088}$ \\
1237661381695570161 & 138.70417 & 31.01759 & 18.881$\pm$0.012 & 0.00055$\pm$0.00001 & 1.742$\pm$0.043 & 7.267$_{-0.043}^{+0.043}$ \\
1237664093439983949 & 138.70486 & 30.26603 & 19.464$\pm$0.026 & 0.00063$\pm$0.00006 & 1.490$\pm$0.025 & 7.034$_{-0.023}^{+0.143}$ \\
1237664668965601602 & 138.70562 & 30.35840 & 19.943$\pm$0.051 & 0.32838$\pm$0.00013 & 1.821$\pm$0.067 & 11.024$_{-0.055}^{+0.054}$ \\
1237664093976985829 & 138.70606 & 30.60060 & 19.234$\pm$0.031 & 0.32198$\pm$0.00005 & 1.192$\pm$0.016 & 10.260$_{-0.076}^{+0.080}$ \\
1237664093439984063 & 138.70617 & 30.13450 & 19.442$\pm$0.043 & 0.26368$\pm$0.00004 & 1.210$\pm$0.036 & 9.984$_{-0.137}^{+0.143}$ \\
\nodata & \nodata &\nodata &\nodata &\nodata &\nodata &\nodata \\
\enddata
\tablenotetext{a}{The survey limiting magnitude is the (Galactic extinction corrected) SDSS Petrosian magnitude ($r_\mathrm{petro}^{lim}=20.75$). The table includes targets with measured stellar masses and D$_n4000$.}
\tablecomments{This table is available in its entirety in  machine-readable form in the online journal. A portion is shown here for guidance regarding its form and content.}
\end{deluxetable*}

The F2 Hectospec data reported by \citet{Geller14} were reduced with an IRAF based pipeline including the RVSAO cross-correlation package \citep{Kurtz98}. We have re-reduced these data with the current pipeline, HSRED v2.0. There is essentially no difference between the relevant measurements, i.e., redshifts or $\dn$, returned by the two pipelines. We report the quantities from the new reduction for consistency with the current on-line MMT database.

Here, we rebuild a redshift survey catalog in the F2 field based on SDSS Data Release DR16 photometry \citep{Ahumada20}. The homogeneous photometry from SDSS enables direct comparison with the local galaxies we select from SDSS. This approach also removes the regions excised from the SHELS catalog as a result of saturation in the DLS photometry. 

From SDSS DR16, we select galaxies in the F2 field with $probPSF = 0$, where $probPSF$ indicates the probability that the object is a star. We then cross-match the SDSS DR16 spectroscopy with the MMT/Hectospec observations in F2. The SDSS DR16 spectroscopy includes 477 and 552 redshifts from SDSS and BOSS, respectively. There are 10037 unique redshifts from the MMT/Hectospec data including 1017 overlaps with SDSS/BOSS. In  cases of overlap, we use the Hectospec redshift. 

\begin{figure*}[h!]
\centering
\includegraphics[width=0.9\textwidth]{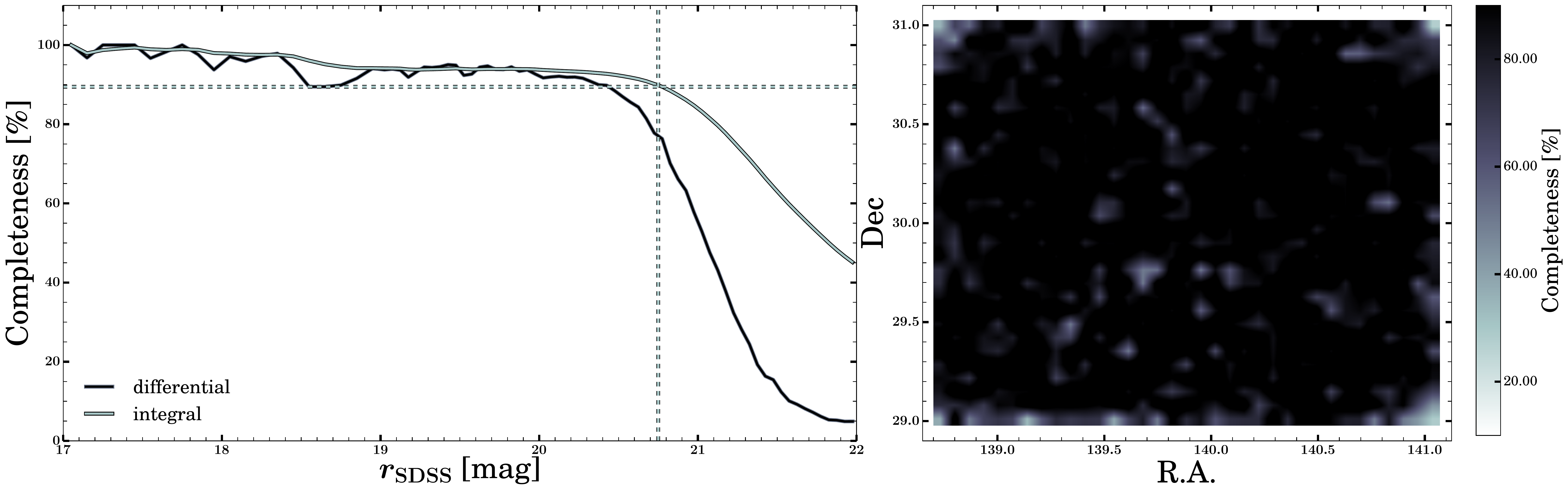}
\caption{{\it Left:} Integral and differential spectroscopic completeness of the SHELS F2 survey as a function of apparent SDSS $r-$band magnitude. The vertical dashed line denotes the magnitude $r_{SDSS}=20.75$ where the integral completeness drops below $90\%$ (dashed horizontal line). {\it Right:} Uniformity of the spatial distribution of the SHELS F2 integral completeness for $17<r_{SDSS}<20.75$.}
\label{fig2}
\end{figure*}

The final catalog (Table \ref{Table2}) is $90\%$ complete to an SDSS Petrosian magnitude $r_{petro} = 20.75$. The catalog covers 4.2 deg$^{2}$, a somewhat larger footprint than the one in \citet{Geller14}. Table \ref{Table2} lists the SDSS Object ID, R.A., Dec, the $r_{petro}$ magnitude, the redshift $z$, the spectral index D$_n$4000, and the stellar mass $\log(M_*/M_\odot$). 

Figure \ref{fig2} shows the differential and integral completeness of the spectroscopic survey as a function of $r-$band magnitude. It also shows the uniformity of the completeness map over the entire F2 field. The catalog includes a total of 11,066 unique redshifts of galaxies brighter than $r_{petro} = 20.75$. 

We review the redshifts in Section \ref{redshifts}. In Section \ref{D4000} we discuss $\dn$ and its use in defining the quiescent population we analyze here. We derive central stellar velocity dispersions in Section \ref{sigma}. In Section \ref{sizes} we describe the measurement of galaxy sizes based on Subaru data. We review the computation of stellar masses in Section \ref{stellarmass}. In Section \ref{SFdmass}, we compute dynamical masses and compare them with SDSS estimates for galaxies observed in both SHELS F2 and SDSS.

\subsubsection{Hectospec Redshifts} \label{redshifts}

In the SHELS F2 field a set of repeat measurements for 1651 unique objects provides the internal estimate (normalized by (1+z)) of the error in the redshift; for emission line objects the internal error is $24~\kms$ and for absorption line objects the internal error is $48~\kms$. In their Hectospec survey of the COSMOS field based on the same observing protocols followed for SHELS, \citet{Damjanov19} obtain similar internal errors for emission and absorption line redshifts, $26~\kms$ and $42~\kms$, respectively. These internal errors are essentially identical to the XCSAO estimate for emission line objects but they exceed it by about 50\% for absorption line spectra.

HectoMAP DR1 \citep{Sohn21}, a larger Hectospec survey, uses the same protocols and reduces the data in the same way as SHELS F2. HectoMAP provides a larger sample for evaluation of the external error. The HectoMAP DR1 includes 129 and 852 observations that overlap SDSS and BOSS, respectively. The mean and standard deviation between redshifts in the total overlapping SDSS/BOSS samples are $39~\kms$ and $44~\kms$, respectively. In other words, Hectospec redshifts exceed those from SDSS/BOSS by a small systematic offset of $39~\kms$. This offset is comparable with the $1\sigma$ standard deviation, $44~\kms$. The mean redshift offset between Hectospec and SDSS/BOSS is insensitive to the S/N of the Hectospec spectra but the external error increases with decreasing S/N as expected. The small offset between SHELS Hectospec and SDSS/BOSS redshifts has no impact on the analysis in this paper.

\subsection{The Quiescent Population}\label{D4000}

\begin{deluxetable*}{lccccc}
\tablecaption{SHELS F2 Quiescent Sample}
\label{Table3}
\tablecolumns{2}
\tabletypesize{\small}
\tablewidth{0pt}
\tablehead{\colhead{SDSS ID} & \colhead{$R_{e}^\tablenotemark{a}$} & \colhead{$b/a^\tablenotemark{b}$}& \colhead{$n$\tablenotemark{c}} & \colhead{$\sigma_*$\tablenotemark{d}}& \colhead{Mass limited sample\tablenotemark{e}}\\
\colhead{} & \colhead{[kpc]} & \colhead{}& \colhead{} & \colhead{[km s$^{-1}$]} & \colhead{}}
\startdata
1237664093976986160 & 4.827$\pm$0.531 & 0.7674$\pm$0.0034 & 1.271$\pm$0.013 & 197.090$\pm$48.311 & 1 \\
1237664669502538244 & 2.083$\pm$0.229 & 0.9148$\pm$0.0059 & 5.192$\pm$0.052 & 179.975$\pm$32.844 & 1 \\
1237664668965601456 & 2.211$\pm$0.243 & 0.7116$\pm$0.0030 & 5.136$\pm$0.026 & 171.180$\pm$19.556 & 1 \\
1237664093440049671 & 7.436$\pm$0.818 & 0.4303$\pm$0.0068 & 4.015$\pm$0.113 & 109.876$\pm$21.987 & 0 \\
1237664092903047907 & 3.417$\pm$0.376 & 0.8258$\pm$0.0070 & 4.547$\pm$0.096 & 150.990$\pm$21.722 & 1 \\
1237664093439984551 & 1.144$\pm$0.126 & 0.1649$\pm$0.0071 & 1.421$\pm$0.025 & 91.535$\pm$26.818 & 1 \\
1237664093976985631 & 3.336$\pm$0.367 & 0.1598$\pm$0.0006 & 1.627$\pm$0.012 & 116.677$\pm$39.017 & 1 \\
1237664667891663387 & 2.237$\pm$0.246 & 0.6761$\pm$0.0039 & 5.644$\pm$0.120 & 194.741$\pm$16.946 & 1 \\
1237664668428665299 & 14.227$\pm$1.565 & 0.9175$\pm$0.0032 & 5.900$\pm$0.043 & 217.901$\pm$18.349 & 1 \\
1237664668428665211 & 3.658$\pm$0.402 & 0.9834$\pm$0.0098 & 5.060$\pm$0.118 & 230.954$\pm$19.669 & 1 \\
\nodata & \nodata &\nodata &\nodata &\nodata &\nodata \\
\enddata
\tablenotetext{a}{\ Circularized half-light (effective) radius. We include only galaxies with $R_e>1$~kpc.}
\tablenotetext{b}{\ Galaxy axis ratio  used to calculate $R_e$ based on the galaxy major axis size $R_a$}
\tablenotetext{c}{\ S\'{e}rsic index}
\tablenotetext{d}{\ Aperture corrected velocity dispersion within 3~kpc}
\tablenotetext{e}{\ 1 denotes inclusion in the stellar mass limited sample}
\tablecomments{This table is available in its entirety in  machine-readable form in the online journal. A portion is shown here for guidance regarding its form and content.}
\end{deluxetable*}

The stellar population age sensitive spectral index, $\dn$, is the flux ratio between two spectral windows (3850 - 3950 \AA\ and 4000 - 4100 \AA) near the 4000~\AA\ break \citep{Balogh99}. \citet{Fabricant08} used repeat Hectospec spectra to show that the internal error in $\dn$ is 1.05 times the value of the index. For 358 spectra in common with SDSS DR6 \citep{SDSSDR6}, the median ratio of the $\dn$ index between independent SDSS and Hectospec measurements is 1. The median redshift of the \citet{Fabricant08} comparison sample is 0.13.

\begin{figure*}[!ht]
\centering
\includegraphics[width=0.9\textwidth]{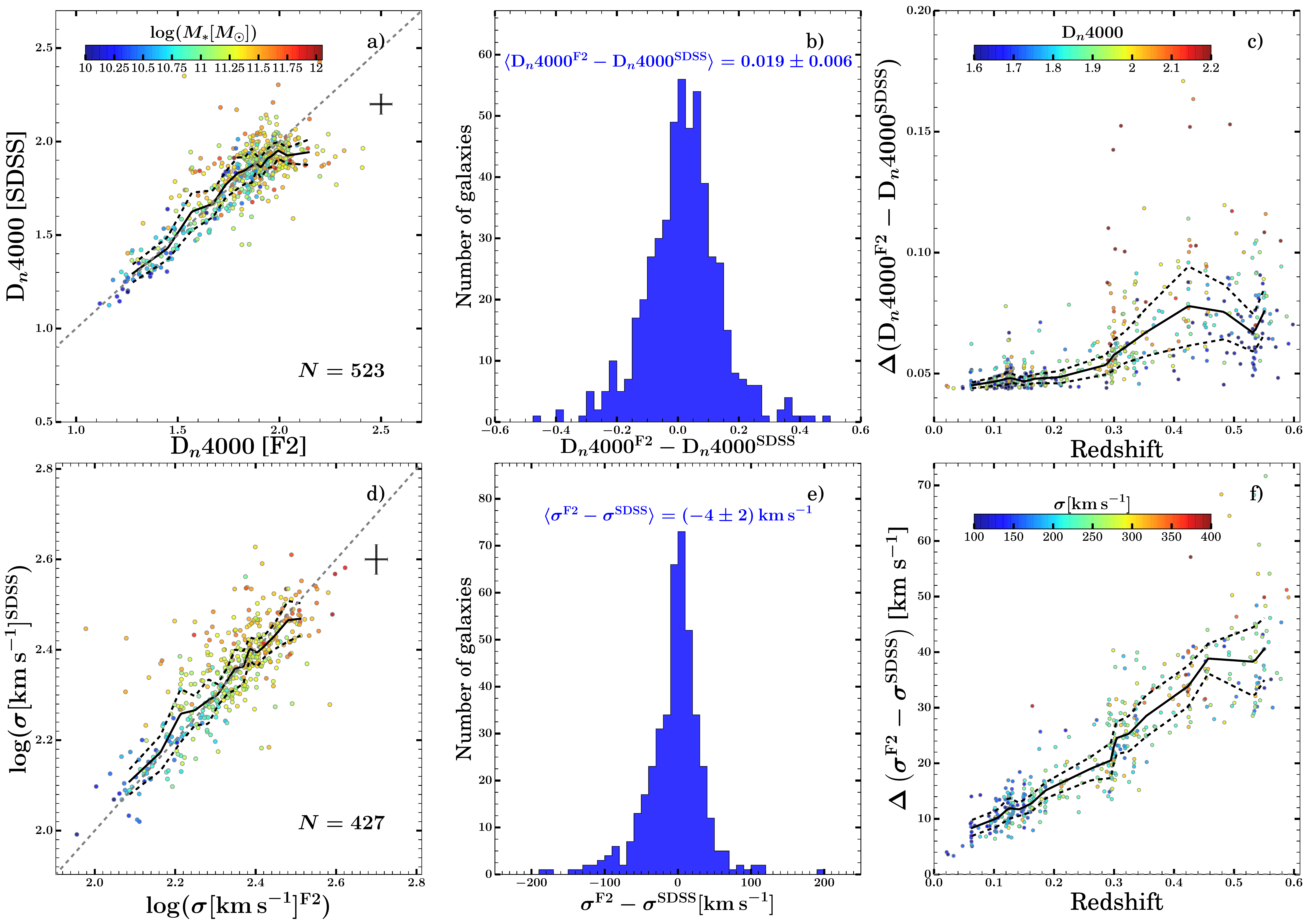}
\caption{{\it Upper panels:} $\dn$ measurements for a subset of SHELS F2 galaxies with high quality MMT/Hectospec {\it and} SDSS/BOSS spectra. The three panels show: a) direct comparison between the two independent measurements (color coded by stellar mass; the single set of error bars  correspond to the combined typical uncertainty in two measurements), b) distribution of differences between SDSS and Hectospec measurements showing that the offset (mean difference) is small and consistent with zero within the $1\sigma$ error, and c) total (propagated) error in the $\dn$ difference as a function of redshift (color coded by the $\dn$ from the MMT/Hectospec spectra). In panel a) dashed lines outline the region: $\rm{median}\left(\rm{D}_n4000\, \rm{[SDSS]}\right)\pm\rm{median}\left|\rm{D}_n4000\, \rm{[SDSS]}-\rm{median}\left(\rm{D}_n4000\, \rm{[SDSS]}\right)\right|$  in equally populated bins of the $\dn$ Hectospec measurements. In panel c) dashed lines outline the region: $\rm{median}\left(\Delta\left(\rm{D}_n4000\right)\right)\pm\rm{median}\left|\Delta\left(\rm{D}_n4000\right)-\rm{median}\left(\rm\Delta\left(\rm{D}_n4000\right)\right)\right|$  in equally populated bins of the $\dn$ Hectospec measurements. {\it Lower panels:} Comparison of SDSS and Hectospec velocity dispersion measurements for the same subset of SHELS F2 galaxies. The measurements are corrected to the same physical aperture, 3 kpc. }
\label{fig3}
\end{figure*}

Extending the analysis of \citet{Fabricant08} to F2, the upper panels of Figure \ref{fig3} compares 523 $S/N > 5$ overlapping SDSS/BOSS \citep{Abazajian2009} and Hectospec measurements of the $\dn$ index. The left panel compares the indices from SDSS/BOSS and SHELS F2 color-coded by stellar mass. The measurements scatter around the one-to-one relation. The solid line connects median $\dn$ values in 16 equally populated mass bins and the dashed curves indicate the $\pm\rm{MAD}$ range, where $\mathrm{MAD}=\mathrm{median}\left(\left|{\mathrm D}_n4000_i-\mathrm{median}\left(\mathrm{D}_n4000\right)\right|\right)$ is the median absolute deviation among the $D_{n}4000$ measurements, a robust measure of variability in each mass bin \citep{Feigelson2012}. 

The central panel of Figure \ref{fig3} shows a histogram of differences between the two independent measures of $\dn$ that reinforces the agreement; the mean difference is 0 within the 1$\sigma$ error. Finally the right-hand panel explores the difference between the $\dn$ measures as a function of redshift. As demonstrated by \citet{Fabricant08} the difference is small at low redshift but it increases to $0.04 - 0.08$ at redshift $z \gtrsim 0.3$. At these larger redshifts the relative scatter also increases.

\citet{Damjanov18}, \citet{Damjanov19}, \citet{Zahid16}, and \citet{Zahid17} use $\dn$ to identify the quiescent population in the COSMOS and F2 fields. They define galaxies with $\dn > 1.5$ as quiescent. \citet[][Section 4.1.]{Damjanov18} make a detailed comparison between $\dn$ and  $UVJ$ color selection. They demonstrate that 90\% of the spectroscopically selected $D_n$4000$ > 1.5$ population in the COSMOS field correspond with systems labeled as quiescent based on their position in the $UVJ$ color-color diagram \citep{williams2009}. \citet{Damjanov19} refine this analysis to focus on the selection boundary. Even in the D$_n$4000 range $1.5-1.6$, 70\% of the spectroscopically selected quiescent objects overlap the morphological classification. The hCOSMOS survey, the basis for the morphological comparison, covers a redshift range similar to the SHELS F2 survey we analyze here. \citet{Damjanov19} argue that a more conservative limit, $\dn > 1.6$ does not affect their analysis of the evolution of the quiescent population. 

Because of the subtle increase in the external error in $\dn$ as a function of redshift (Figure \ref{fig3}), we test the difference between samples with different selection criteria under the assumption that all galaxies with $\dn > 1.55$  are truly quiescent \citep{Kauffmann03}. We perform a suite of simple simulations by sampling 3266 $0.4 < z <0.6$ galaxies from our parent sample 1000 times. In each of the simulated samples we draw individual $\dn$ values by altering the measurement with a random error drawn from a Gaussian distribution centered at zero and with a scale corresponding to the measurement error. We construct a figure of merit that compares the distribution of the true positive rate ($TPR$, the ratio between the number of true positives and the sum of true positives and false negatives) and the false positive rate ($FPR$, the ratio between the number of false positives and the sum of true negatives and false positives) for the selection based on two $\dn$ cuts ($>1.6$ and $>1.5$). The average $FPR$ spans the same range ($\sim1.5-3\%$) for both selection criteria. In contrast, the average $TPR$ is $\sim96\%$ for $\dn > 1.6$ selection and only $\sim 87\%$ for $\dn > 1.5$ selection. Thus, in recognition  of the larger error at larger redshift, we adopt the more conservative limit $\dn > 1.6$ to define the quiescent population throughout the F2 survey.

\subsubsection{Central Velocity Dispersion}\label{sigma}

We collected velocity dispersion measurements for the 1192 SDSS and BOSS galaxies from the Portsmouth reduction \citep{Thomas13}. Portsmouth reduction velocity dispersions are based on Penalized Pixel-Fitting (pPXF, \citealp{Cappellari04}). The best-fit velocity dispersions compare SDSS spectra with stellar population templates \citep{Maraston11} generated from the MILES stellar library \citep{SanchezBlazquez06}). The median error for SDSS/BOSS galaxies in F2 with $90 \rm\ {km~s^{-1}} < \sigma < 600 \rm\ {km~ s^{-1} }$ is $\sim25$~km~s$^{-1}$. 

We measure central stellar velocity dispersions from Hectospec spectra using the pPXF-based UlySS code (University of Lyon Spectroscopic analysis Software, \citealp{Koleva09}). We use the PEGASE-HR code \citep[e.g.,][]{LeBorgne04} to construct stellar population templates based on the MILES stellar library. ULySS convolves these templates to the Hectospec resolution at varying velocity dispersions, stellar population ages, and metallicities. Then $\chi^{2}$ minimization finds the best-fit age, metallicity, and velocity dispersion. \citet{Fabricant13} show that the error in the velocity dispersion is minimized by deriving the fit within the rest-frame spectral range 4100--5500 \AA. The typical uncertainty in the Hectospec velocity dispersions in the rebuilt F2 sample is $32~\kms$. 

Among the SHELS F2 galaxies with $r<20.75$, D$_n4000 >1.6$ and $R_e >1$~kpc, 3068/3355 (91\%) have a measured central stellar velocity dispersion in the range $90\, \rm{km~ s}^{-1}<\sigma<600\, \rm\ {km~ s}^{-1}$. Among these objects, 427 overlap with the SDSS. The lower panels of Figure \ref{fig3} compare the SDSS/BOSS and MMT velocity dispersion measurements for this subset of 523 overlapping objects we used to compare $\dn$ measurements. 

We compare the velocity dispersions corrected to a fixed physical aperture of 3 kpc. This radius corresponds to the 1.5$^{\prime\prime}$ Hectospec aperture at $z \sim 0.3$. The corrections are very small \citep{Zahid16}. We refer to this aperture corrected central stellar velocity dispersion as $\sigma_*$.

The lower left panel of Figure \ref{fig3} compares the SDSS/BOSS and F2/Hectospec values of $\sigma_*$. The points are again color-coded by stellar mass. Although there are some outliers, the dispersions are generally in agreement: the solid line that connects median values in equally populated mass bins follows the $1:1$ relation and the $\pm\, \rm{MAD}$ range (dashed lines) is narrow. The histogram of dispersion differences in the central panel confirms the agreement; the mean difference is much smaller than the error in an individual $\sigma_*$. The right panel of Figure \ref{fig3} shows  the total error in the difference between the two dispersions as a function of redshift. This error increases as a function of redshift: at $ z\sim 0.5$, the difference between the two velocity dispersion measurements is in the MAD range of $\sim(-50,50)\, \rm{km\, s}^{-1}$ and is thus very similar to the combined average error at the same redshift. This difference is irrelevant for our analysis.

\subsubsection{Subaru Photometry and Sizes}\label{sizes}

We measure galaxy sizes for the galaxies with redshifts in F2 from Subaru Hyper Suprime-Cam (HSC; \citealp{Miyazaki18}) $i-$band images. \citet{Utsumi16} describe the HSC observing procedure and image processing in detail; we summarize the procedure briefly here.

The HSC image, taken in 0.5 - 0.7 arcsec seeing, includes 18 pointings of 240 seconds exposures. These pointings overlap and extend beyond the 4 deg$^{2}$ footprint of the F2 field to yield uniform depth (see Figure 1 in \citealp{Utsumi16}). Galaxy number counts show that the HSC F2 image provides a complete catalog of extended sources to a limiting $i\simeq 25$.

The images are processes using the hscPipe system \citep{Bosch18}, the standard pipeline for the HSC Subaru Strategic Program (SSP). The pipeline facilitates reduction of individual chips, mosaicking, and image stacking. 

As in our earlier exploration of quescent galaxies size and spectroscopic evolution \citep{Damjanov19}, we employ single Sérsic profile models \citep{Sersic68} to estimate the sizes of the SHELS F2 galaxies. \citet{Damjanov19} describe the approach and compare the results with previous size measurements in detail; we briefly summarize the procedure here.

We use the SExtractor software \citep{Bertin96} to measure photometric parameters of galaxies with redshifts in F2. These parameters  include the galaxy half-light radius, ellipticity, and Sérsic index. These parameters are based on two-dimensional (2D) modeling of the galaxy surface brightness profile following a three-step process: (a) in its first run, SExtractor provides a catalog of sources that includes star–galaxy separation; (b) the PSFex software \citep{Bertin11} combines point sources from the initial SExtractor catalog to construct a set of spatially varying point-spread functions (PSFs) that are used as input parameters for (c) the second SExtractor run to provide a catalog with morphological parameters for all detected sources.

Within the SPHEROID model, the SExtractor fitting procedure provides Sérsic profile parameters and the best-fit model axial ratio $b/a$ (SExtractor model parameter SPHEROID$\_$ASPECT$\_$WORLD). We use the angular diameter distance corresponding to the spectroscopic redshift of each galaxy to translate the galaxy angular size $\theta_a$ (SExtractor model parameter SPHEROID$\_$REFF$\_$WORLD) into the major axis radius $R_a$ in kiloparsecs \citep[e.g.,][and references therein]{Hogg99}. 

Many studies are based on the circularized half-light radius; others \citep[e.g.,][]{Belli17} use the semi-major axis. We use the circularized half-light radius, ${R}_{e}={R}_{a}\times \sqrt{b/a}$. The mean circularized half-light radius is 4.5 kpc, and the median is 3.6 kpc. We have checked that neither the slopes of the related scaling relations nor the evolutionary trends that we derive below depend on this choice. We also compute the stellar velocity dispersion within R$_e$, $\sigma_e$. The ratio $\sigma_e/\sigma_*$ ranges from 0.91 to 1.037 with a median of 0.994. We use $\sigma_e$ throughout the following analysis.

\subsubsection{Stellar Masses}\label{stellarmass}

We calculate stellar masses from the SDSS DR16 $ugriz$ model magnitudes corrected for foreground extinction \citep[e.g.,][] {Geller14, Zahid16}. We fit the observed spectral energy distribution (SED) with the Le PHARE fitting code\footnote[1]{http//www.cfht.hawaii.edu/~arnouts/LEPHARE/lephare.html}. We incorporate the stellar population synthesis models of \citet{Bruzual03} and we assume a universal Chabrier initial mass function \citep[IMF,][]{Chabrier03}. The models incorporate two metallicities and exponentially declining star formation rates. The e-folding time for star formation ranges from 0.1 to 30 Gyr. Model SEDs include the internal extinction range $E(B-V) = 0 - 0.6$ based on the \citet{Calzetti00} extinction law. The population age ranges from 0.01 to 13 Gyr. We normalize each SED to solar luminosity. The ratio between the observed and synthetic SED is the stellar mass. The median of the distribution of best fit stellar masses is our estimate of the stellar mass. 

Stellar masses carry a systematic uncertainty of $\sim 0.3$ dex. This uncertainty affects the normalization of the scaling relations we discuss, but it should not affect the slopes. These systematics can produce a shift in the median stellar mass for a particular sample, but they do not change the shape of the distributions or the relative medians for different samples.

Table \ref{Table3}, the first of two data summary tables, lists the measurements of physical properties for galaxies in the F2 quiescent sample. The table includes the SDSS Object ID, the circularized effective radius $R_e$, the axial ratio $b/a$, the Sersic index $n$, the stellar velocity dispersion $\sigma_*$ within a fiducial 3 kpc aperture, and a flag indicating inclusion in the mass limited sample. Based on MMT/Hectopsec data, we provide the first velocity dispersion measurements for 2985 out of 3068 sources with $r<20.75$, D$_n4000 > 1.6$, and $R_e>1$~kpc. 

Table \ref{Table4} lists the number of galaxies in the SDSS, SHELS F2, and high redshift samples that satisfy various selection criteria. The subscripts in the left-hand column specify the relevant requirements; a subscript without a limit simply indicates that the data must include a measurement for that quantity (i.e., even though all of the galaxies have spectroscopy, a few lack a value of D$_n$4000 and or $M_*$ as a result of poor spectroscopy and/or photometry). We note that the high redshift samples are not magnitude limited and they do not have a measured D$_n$4000.

\begin{deluxetable*}{lccc}
\tabletypesize{\small}
\tablecaption{Number of galaxies in the $0<z<2.5$ samples}
\label{Table4}
\tablewidth{7in}
\tablehead{
\colhead{Sample} & \multicolumn{2}{c}{Number of Galaxies}\\
\colhead{} & \colhead{F2} & \colhead{SDSS} & \colhead{High-$z$}
}
\colnumbers
\startdata 
{SDSS $r-$band limit [mag]} & 20.75 & 17.78 &\nodata \tablenotemark{a}\\
$z$ & (0.1, 0.6) & (0.05, 0.07) & (0.8,3.8)\\
$N_\mathrm{spec}$ & 10848 &  88960 & 126\\
$N_{\mathrm{spec},\, \mathrm{D}_n4000}$ & 10821 & 88958 &\nodata\tablenotemark{b}\\
$N_{\mathrm{D}_n4000>1.6}$ & 3831 & 47363 & \nodata\tablenotemark{b}\\
$N_{\mathrm{D}_n4000>1.6,\, M_\ast}$ & 3812 & 44268 & \nodata\tablenotemark{b}\\
$N_{\mathrm{D}_n4000>1.6,\, M_\ast,\, \sigma_\ast}$ & 3465 & 33843& \nodata\tablenotemark{b}\\
$N_{\mathrm{D}_n4000>1.6,\, M_\ast,\, \sigma_\ast, R_e>1\, \mathrm{kpc}}$ & 3046 & 32299&\nodata\tablenotemark{b}\\
$N_\mathrm{mass\, limited\, sample}$ & 2906 & 32097& \nodata\tablenotemark{b}\\
$N_{10.7<\log(M_\ast/M_\sun)<11.5}\tablenotemark{c}$ & 2040 & 8178 & 99\\
\enddata

\tablenotetext{a}{\ The high-z samples are not magnitude limited.}
\tablenotetext{b}{\ The number of objects is constant in the high redshift sample because D$_n$4000 is not available for these data.}
\tablenotetext{c}{\ The number of objects within the stellar mass range we use for the evolutionary trends in the right panels of Figures~\ref{fig6},~\ref{fig7},~and~\ref{fig9}.} 
\end{deluxetable*}

\begin{figure*}[!ht]
    \centering
    \includegraphics[width=0.9\textwidth]{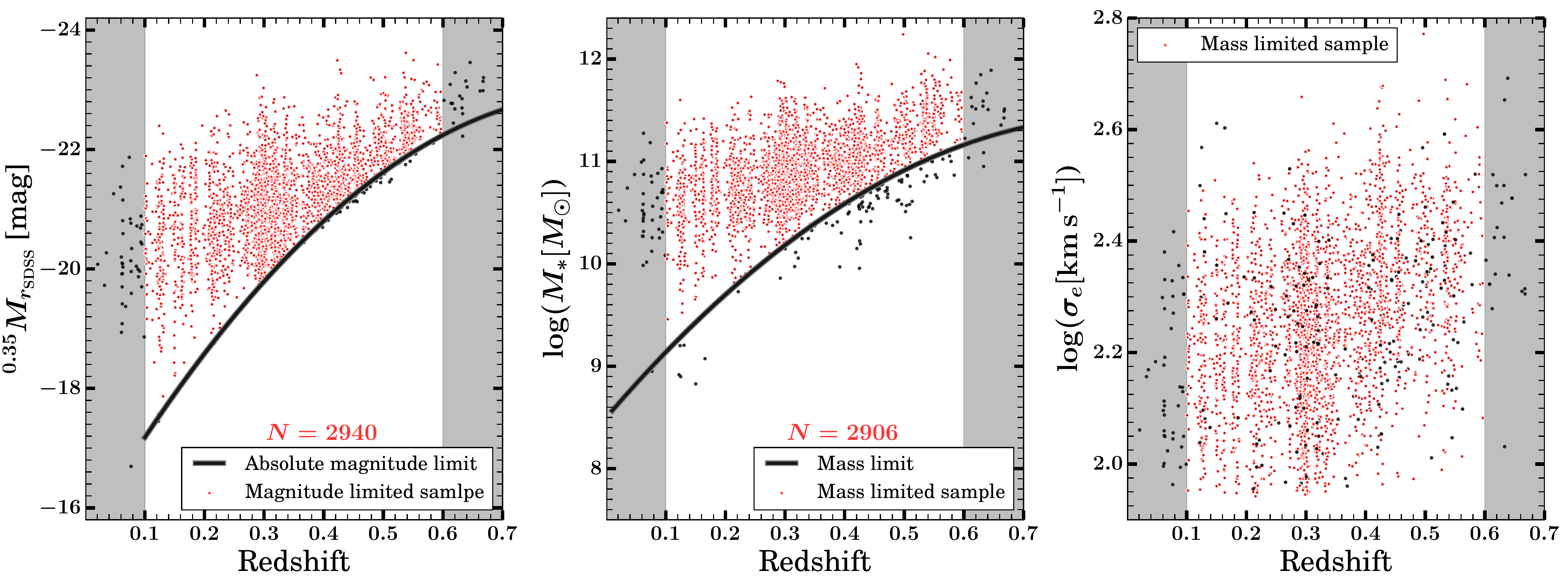}
    \caption{Absolute $r-$band magnitude ({\it left} panel), stellar mass ({\it center} panel), and velocity dispersion ({\it right} panel) versus redshift for the $0.1 \leq z \leq 0.6$ SHELS F2 quiescent galaxy sample. The absolute magnitude limit (black curve in the left panel) corresponds to the K-corrected $r_{petro} = 20.75$ magnitude limit of the parent survey. The stellar mass limit (black curve in the central panel) is based on the average mass-to-light ratio for SHELS F2 galaxies at each redshift. Black points indicate the $r_{petro} = 20.75$ intermediate redshift quiescent sample and red points denote the absolute magnitude (left panel) and mass (central and right panel) complete subsamples. The legends give the total number of quiescent SHELS F2 systems in the absolute magnitude and mass limited samples. We use only the unshaded redshift range for further analysis.}
    \label{fig4}
\end{figure*}

The left panel of Figure \ref{fig4} shows the K-corrected Petrosian $r-$band absolute magnitudes for the SHELS F2 quiescent sample as a function of redshift. The line shows the magnitude limit corresponding to the K-corrected $r_{petro} = 20.75$ apparent magnitude limit of the F2 survey. The central panel shows the stellar mass as a function of redshift along with the completeness limit derived from applying the mean mass-to-light ratio for the sample as a function of redshift. The right panel shows the central velocity dispersion as a function of redshift. We omit the gray-shaded regions from further analysis because the sampling in these redshift ranges is sparse. Similarly to the SDSS sample (Figure~\ref{fig1}), at $0.1<z<0.6$ a negligible fraction of SHELS F2 quiescent systems from the absolute magnitude limited sample lie below the stellar mass limit (Figure~\ref{fig4}, central panel, black points in the non-shaded region). Furthermore, these galaxies have uniformly distributed velocity dispersion measurements (Figure~\ref{fig4}, right panel, black points in the non-shaded region), just like their SDSS counterparts. Red points in the central and right-hand panels indicate the subset of the objects that constitute a complete mass limited sample.

\subsection{Distributions of Stellar Mass, Radius, and Central Stellar Velocity Dispersion}
\label{allsamples}

The galaxy samples we include cover very different redshift ranges. The distributions of $M_\ast$, $R_e$, and $\sigma_e$ overlap at different redshifts but do not coincide. Table \ref{Table1} details the sources for the physical parameters of galaxies in the SDSS and F2 samples. 

We explore existing high redshift samples in two comparably populated redshift bins (Table \ref{Table4}). In the redshift range $0.8\lesssim z \lesssim2$, the data are from \citet{vdS13}, who include a compilation of high-z studies with FAST-derived  stellar masses \citep[][53/73]{Kriek2009}, pPXF-derived velocity dispersions \citep{Cappellari04,Cappellari2017}, and GALFIT structural parameters \citep{Peng2010}. For the higher redshift range ($1.5\lesssim z\lesssim 3.8$), we include data from \citet{Belli17}, \citet{Stockmann2020}, \citet{Mendel2020}, and \citet{Esdaile2021}. Their data include stellar masses derived using either FAST or Bayesian-based approaches, velocity dispersions measured using either pPXF or simultaneous fitting of FSPS models \citep{Conroy2009,Conroy2010} to spectra and multi-band photometry, and GALFIT structural parameters. 

\begin{figure*}[h!]
\centering
\includegraphics[width=0.75\textwidth]{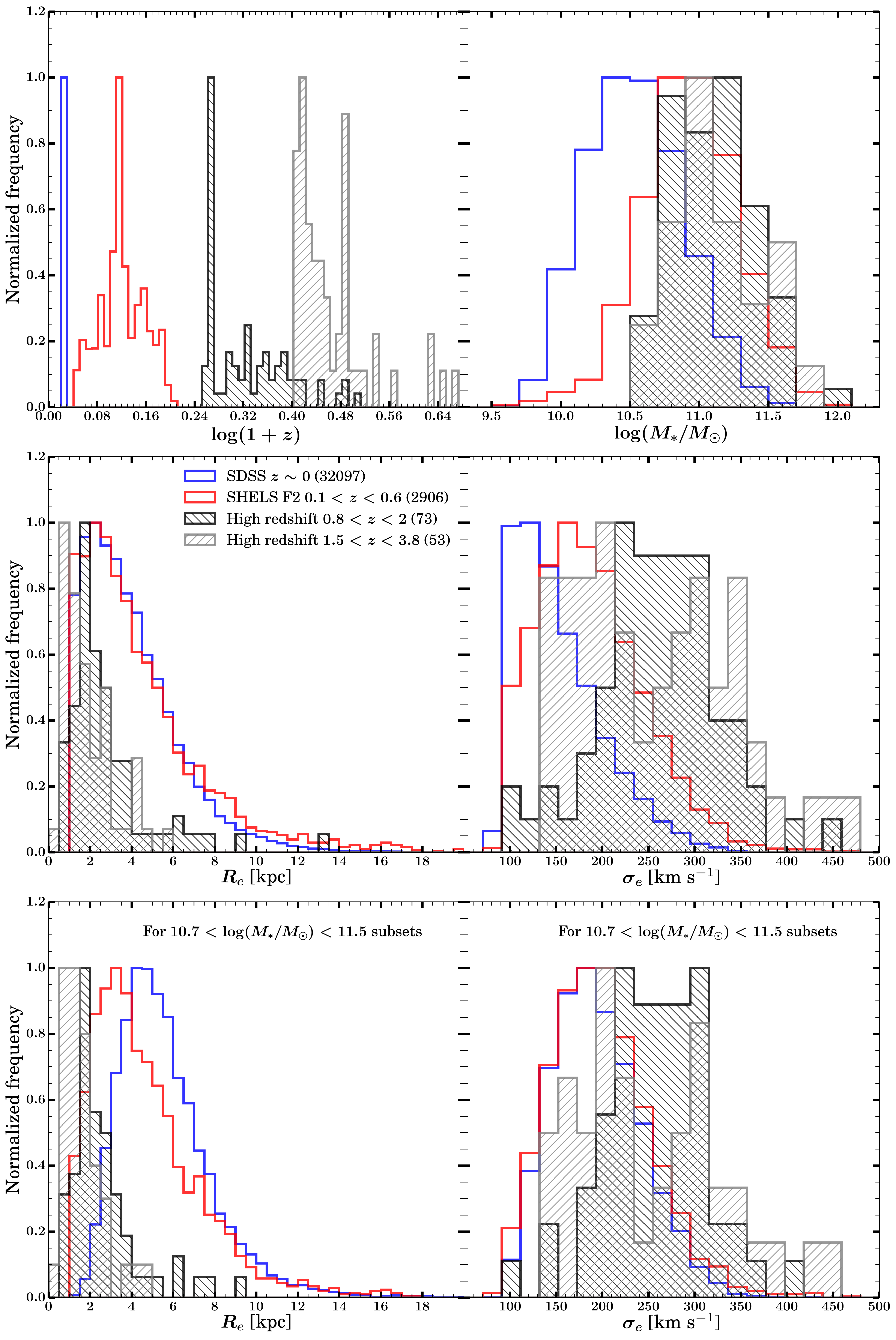}
\caption{Distribution of redshift, stellar mass, effective (circularized) radius, and velocity dispersion for samples of quiescent galaxies spanning the redshift range $0 < z <3.8$ (first two rows). The histograms are normalized to the most populated bin in each histogram. Only the SDSS ($0.05 < z < 0.07$) (blue) and SHELS F2 ($0.1 < z < 0.6$) (red) are stellar mass complete. The high-$z$ comparison samples are from \citet{vdS13, vdS15} (black) and from \citet {Belli17,Stockmann2020,Mendel2020,Esdaile2021} (gray). We use the relatively narrow mass range where all samples (from $z\sim0$ to $z\sim3.5$) overlap, $10.7 < \log (M_{*} / M_{\odot}) < 11.5$, to trace evolution in velocity dispersion (Figure \ref{fig6}), size (Figure \ref{fig7}), and dynamical mass (Figure \ref{fig9}). The two lower panels show the effective (circularized) radius and velocity dispersion distributions limited to the mass range $10.7 < \log (M_{*} / M_{\odot}) < 11.5$ only. }
\label{fig5}
\end{figure*}

Figure \ref{fig5} shows distributions of the redshift (first row left), stellar mass (first row right), effective radius (second row left), velocity dispersion (second row right). Figure \ref{fig5} also shows (third row) effective radius and velocity dispersion distributions for $10.7<\log(M_\ast/M_\sun)<11.5$ subsamples throughout the redshift range. The four histograms covering different redshift ranges correspond to SDSS (blue), SHELS F2 (red), and the two high redshift samples: \citet{vdS13} (black); \citet{Belli17}, \citet{Stockmann2020}, \citet{Mendel2020}, and \citet{Esdaile2021} (gray). Both the SDSS and SHELS F2 samples are complete in stellar mass at each redshift; these samples are denoted by red points in the central panels of Figures \ref{fig1} and \ref{fig4}, respectively. None of the high redshift samples are complete in stellar mass; we simply take the aggregate sample of available objects as published for a basic comparison with the lower redshift data. 

The stellar mass distributions of the two highest redshift samples (black and gray) are similar but they are limited to masses $\gtrsim 10^{10.5}\, M_\odot$. For the highest redshift subsample (gray) the peak effective radius, $R_e \sim 0.5$ kpc, is smaller than the $R_e \sim 2$ kpc for $0.8\lesssim z\lesssim 2$ subsample. A Kolmogorov-Smirnov (KS) test\footnote{\url{https://docs.scipy.org/doc/scipy/reference/generated/scipy.stats.ks_2samp.html}} shows that the distributions of velocity dispersions for the two highest redshift samples are unlikely to be drawn from the same parent distribution ($p=0.05$). The highest redshift sample includes a significant number of objects with $\sigma_e \lesssim 200$ km s$^{-1}$. In the $0.8<z<2.5$ sample, many of the objects are members of a cluster of galaxies at $z \sim 0.8$, a possible selection effect. For example, \citet{Sohn2020} show that high dispersion objects tend to be overabundant in clusters.

Figure \ref{fig5} shows that, because of the large redshift range we probe, the samples overlap only in a relatively narrow stellar mass range.  We use the overlap range $10.7 < \log (M_{*} / M_{\odot}) < 11.5$ to trace evolution in size and velocity dispersion over the full redshift range. The bottom left panel of Figure \ref{fig5} shows  that the mode of the size distributions shifts to smaller sizes as the redshift increases. In contrast, for the two mass-limited susbamples at $z<0.6$, the velocity dispersion distributions are identical (blue and red histogram of Figure \ref{fig5}, bottom right). However, the subsamples at $z\gtrsim0.8$ still appear to differ (gray and black histograms of Figure \ref{fig5}, bottom right) in the same sense as the full high redshift samples
(Figure \ref{fig5}, middle right panel).

\section{Scaling Relations and Redshift Evolution }
\label{scaleevolution}

Scaling relations between photometric and spectroscopic properties of quiescent galaxies are a fundamental basis for measuring and understanding their evolution. Here we revisit the scaling relations between velocity dispersion and stellar mass and between size and stellar mass. We then use the samples described in Figure \ref{fig5} to constrain the redshift evolution of quiescent galaxy size, velocity dispersion, and dynamical mass at fixed stellar mass. We explore these issues for the limited stellar mass range (10.7 $ < \rm{log} (M_*/M_\odot) < 11.5$) where the samples overlap (Figure \ref{fig5}).

\citet{Zahid16} use stellar mass limited samples of quiescent galaxies in SHELS F2 and SDSS to explore the relation between central velocity dispersion and stellar mass. They show that for quiescent galaxies with stellar masses $M_* > 10^{9.5}\, M_\odot$ and $z < 0.7$, $\sigma_e* \propto M_*^{0.3}$. This result agrees with other studies based on large low-redshift ($z\lesssim0.2$) samples \citep[e.g.,][]{Gallazzi2006,Hyde09,Shankar2010,Aguerri2012,Napolitano2020}. For example, \citet{Hyde09} analyze quiescent galaxies in the SDSS and find a similar slope. In \citet{Zahid16}, quiescent objects have D$_n4000 > 1.5$. The results do not change for a sample selected with D$_n4000 > 1.6$, the cut we adopt.

\citet{Cannarozzo20} analyze a smaller sample of objects and obtain a redshift independent shallower slope of $\sim 0.18$. This shallower slope probably reflects sampling issues including  the steepening of the relation for stellar masses below the $M_* = 3 \times 10^{10}\, M_\odot$ cutoff applied by \citealp{Cannarozzo20} \citep[see e.g., Figure~11 in][]{Zahid17} and differences in analytic approach. In additional tests of the $M_* -\sigma_e*$ relation, \citet{Zahid16} demonstrate that a slope of $\sim 0.3$ characterizes both distance limited subsamples of the SDSS and the high redshift sample of \citet{Belli2014}.

\begin{figure*}[h!]
\centering
\includegraphics[width=0.9\textwidth]{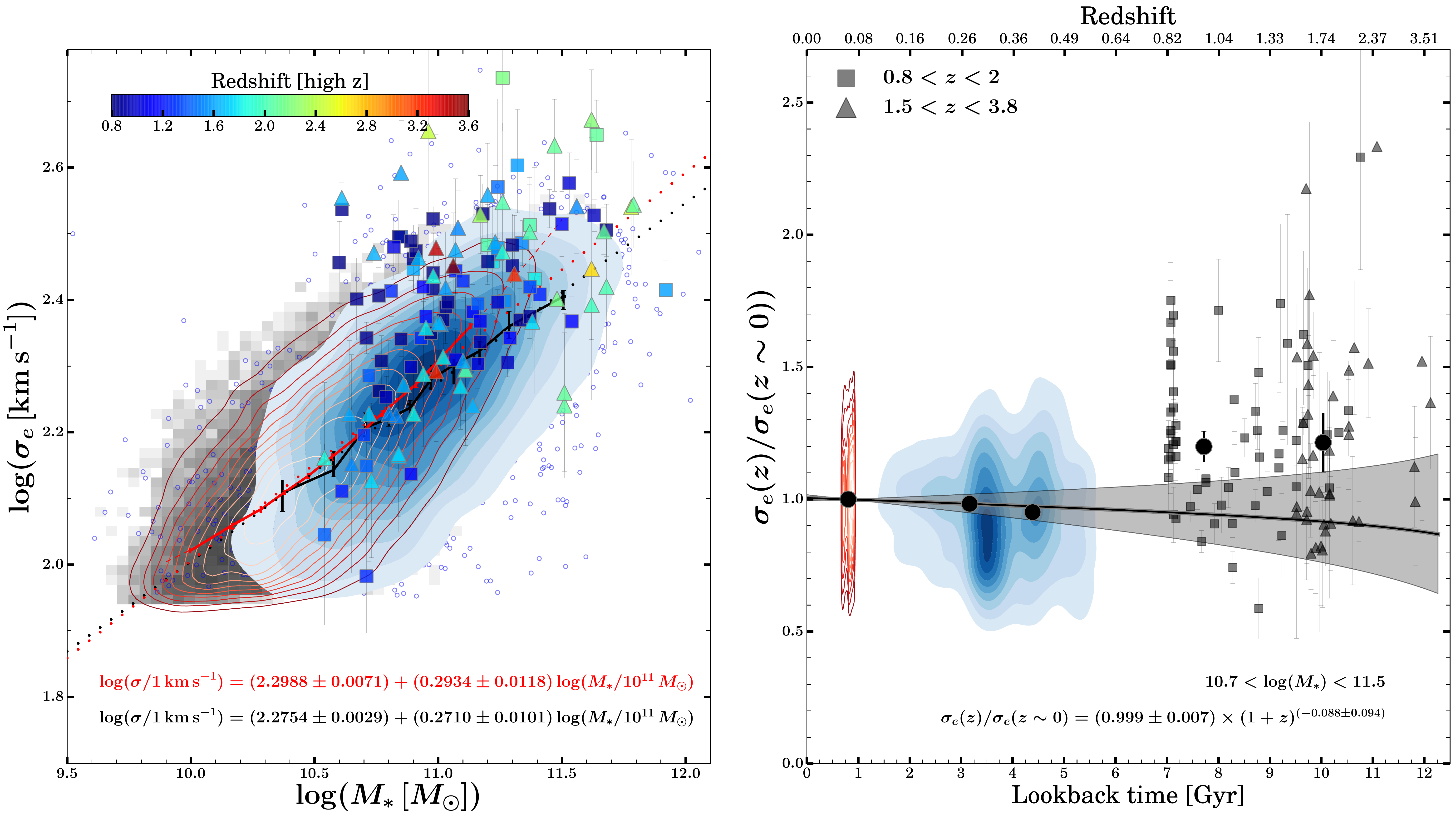}
\caption{(Left) Stellar velocity dispersion within the effective radius vs. stellar mass for three redshift intervals: $0.05<z<0.07$ (SDSS, grey 2D histogram and red contours of constant density), $0.1<z<0.6$ (SHELS F2, blue points and blue  contours), and the high-$z$ samples from Figure~\ref{fig5}, with symbols color-coded by redshift and described in the legend (right panel). The red (black) solid line connects median velocity dispersions for SDSS (SHELS F2) galaxies in equally populated mass bins. The red dashed line denotes the interpolation function (shown in the extrapolation interval for clarity) for the SDSS sample used to calculate $\sigma_e(z)/\sigma_e(z\sim0)$ at greater redshift. The dotted lines and the legend show the best linear fits to the median velocity dispersion - stellar mass relation for the SDSS sample (red) and SHELS F2 sample (black). There is almost no evolution  in either the zero-point or the slope. (Right) Evolution in velocity dispersion at fixed stellar mass over 12~Gyrs of cosmic time (i.e. $z\sim0$ to $z\sim3.5$ on the upper axis). Red (blue) contours show the SDSS (SHELS F2) sample. Solid black circles with error bars represent median ratios in different redshift bins. Squares and triangles show the velocity dispersion ratios for individual galaxies from the tow high redshift bins. The black solid line and shaded region show the best-fit $\sigma_e(z)/\sigma_e(z \sim 0) = \alpha (1+z)^\beta$ relation $\pm$ $2\sigma$ range for an exponent $\beta$ (if $\alpha$=1). The velocity dispersion  shows no significant evolution.}
\label{fig6}
\end{figure*}

The left panel of Figure \ref{fig6} revisits the scaling between $\sigma_e$, the stellar velocity dispersion within the effective radius, and $M_*$ for SDSS and SHELS F2 stellar mass complete samples (Figures~\ref{fig1} and~\ref{fig4}) following \citet{Zahid16}. Figure \ref{fig6} shows the distribution of galaxies in four samples: SDSS (gray histogram, red contours and curve), SHELS F2 (blue points and contours, black curve), and the two  high redshift samples following Figure \ref{fig5}. The red and black curves show the median value of $\sigma_e$ in equal mass bins. The normalization and slope for both samples agree well with the  analysis of \citet{Zahid16}. There is essentially no evolution in this scaling relation in the redshift range covered by the SDSS and SHELS F2 surveys. 

The high redshift samples are not complete in stellar mass, but the data points generally track a relation with a slope similar to the lower redshift samples. These samples miss quiescent objects with $M_\ast\lesssim 3\times10^{10}\, M_\sun$. Furthermore, they generally  include lower velocity dispersion systems (i.e., galaxies that lie below median curves for $z\sim0$ and $0.1<z<0.6$ samples) only in relatively narrow stellar mass bins (at $M_\ast\sim5\times10^{10}\, M_\sun$ and $M_\ast\sim3\times10^{11}\, M_\sun$). These properties of the high redshift samples probably result from  selection issues. 

The right panel of Figure \ref{fig6} shows the limits on the evolution of $\sigma_e$ placed by these  samples. The plot shows the ratio $\sigma_e(z)/\sigma_e(z \sim 0)$ as a function of both redshift, $z$ (upper axis), and lookback time (lower axis). For each object in the sample we compute $\sigma_e(z \sim 0)$ based on the dotted line in the left panel. This line is an extrapolated fit to the median SDSS relation between $\sigma_e$ and $M_*$. 

The red (blue) contours in Figure \ref{fig6} show the SDSS and SHELS F2 data, respectively, and squares and triangles denote the two high redshift samples (black and gray histograms, respectively, from Figure \ref{fig5}). The heavy black points show the median in each redshift bin; the black curve is a fit to these points. Following \citet{vdS13} we fit the function $\sigma_e(z)/\sigma_e(z \sim 0)=\alpha (1+z)^\beta$. We find $\alpha = 0.999 \pm 0.007$, indistinguishable from unity and $\beta = -0.1 \pm 0.1$. The gray region shows the 2$\sigma$ error range for $\alpha = 1$. 

The fit to all of the data implies that there is no significant evolution in $\sigma_e$. For the high redshift samples, the median value for $1.5<z<3.8$ systems is within $\sim 2 \sigma$ of the error range in the fit. For the lower redshift subset ($0.8<z\leq 1.5$), the median $\sigma_e$  is discrepant with the best-fit relation.  \citet{vdS13} find an increase in $\sigma_e$ with decreasing redshift,  $\beta = 0.49 \pm 0.08$, driven by the difference between the median SDSS velocity dispersion combined exclusively with their high redshift data.

There are three important differences between our analysis and those like \citet{vdS13} that combine a high redshift sample with the SDSS. First, we analyze the velocity dispersion as a function of stellar rather than dynamical mass. Unlike the dynamical mass, the stellar mass is independent of the velocity dispersion. Second, we include the SHELS F2 data at intermediate redshift; thus we have a much more restrictive baseline for the fit based on the combined, large SDSS and SHELS F2 samples. In fact, if we remove the SHELS F2 data, we recover a mild increase in $\sigma_e$ with decreasing redshift similar to work that analyzes these data combined with SDSS.  Finally, unlike all of the high redshift samples, both the SDSS and SHELS F2 samples are complete in stellar mass. 

In contrast with the robust evolution in the size of quiescent galaxies that we discuss next, Figure \ref{fig6} suggests that any evolution in velocity dispersion is subtle. Dense sampling of the redshift range between the SDSS coverage and the high redshift samples is thus crucial for constraining the redshift evolution of the quiescent population. Figure \ref{fig6} also underscores the importance of stellar mass complete samples throughout the redshift range.

\begin{figure*}[h!]
\centering
\includegraphics[width=0.9\textwidth]{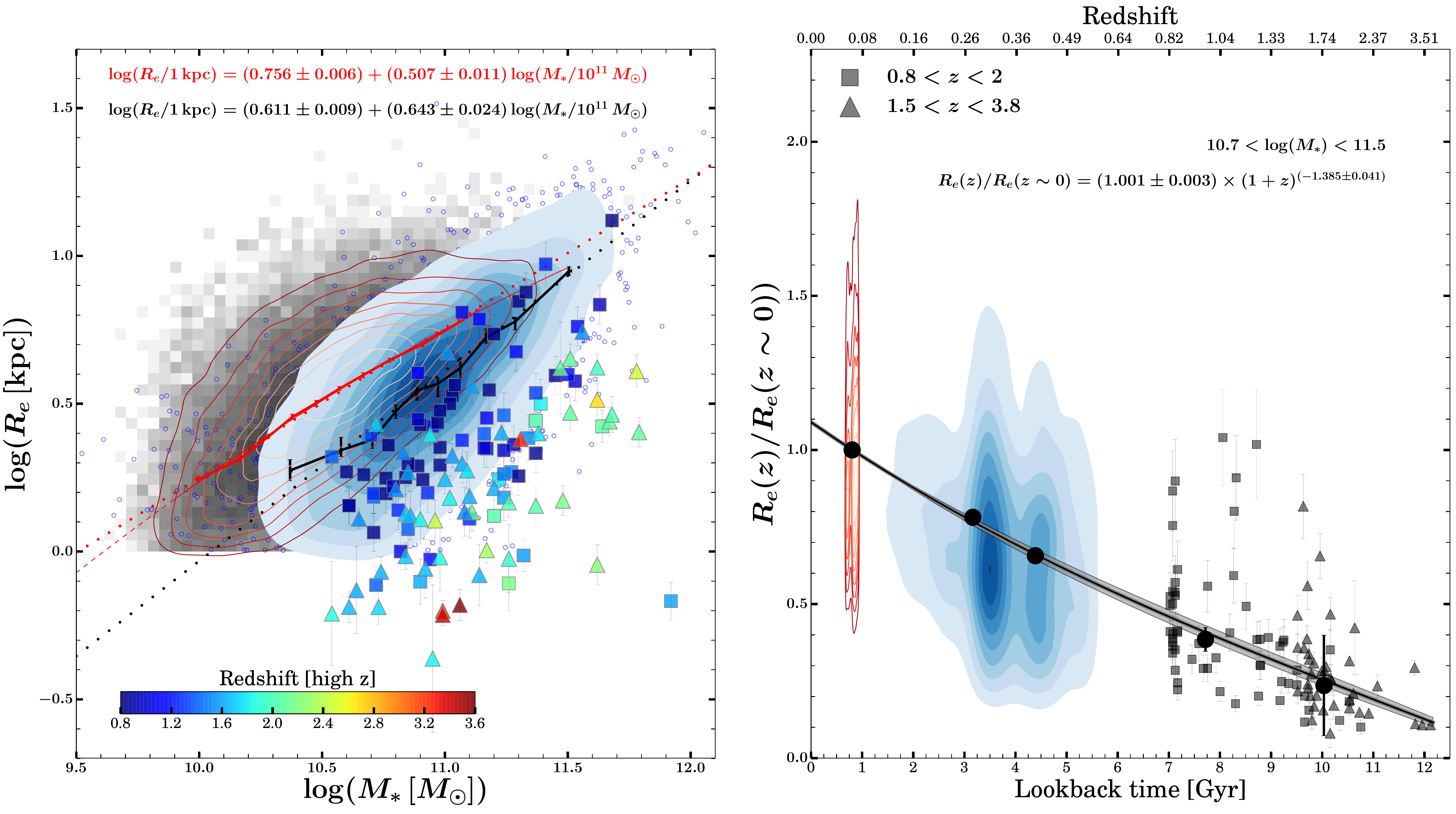}
\caption{(Left) Effective radius (R$_e$) vs. stellar mass for three redshift intervals: $0.05<z<0.07$ (SDSS, grey 2D histogram and red  contours of constant density), $0.1<z<0.6$ (SHELS F2, blue points and  contours), and $0.8<z<3.8$ sample (symbols color-coded by redshift and described in the right-hand panel of Figure~\ref{fig6}). The red (black) solid line connects median R$_e$ values for SDSS (SHELS F2) galaxies in equally populated mass bins. The red dashed line denotes the interpolation function (shown in the extrapolation interval for clarity) for the SDSS sample  used to calculate R$_e$(z)/R$_e$(z$\sim$ 0) at higher redshift. The dotted lines and the legend show the best linear fits to the median size -stellar mass relation for the SDSS sample (red) and the SHELS F2 sample (black). (Right) Evolution in effective radius at fixed stellar mass over 12~Gyrs of cosmic time (i.e., $z\sim0$ to $z\sim3.5$ on the upper axis). Red (blue) contours show  the SDSS (SHELS F2) sample. Solid black circles with error bars represent median ratios in different redshift bins. Squares and triangles  show individual high-$z$ galaxies as in the left-hand panel.  The black solid line and shaded region show the best-fit $\rm{R}_e(z)/\rm{R}_e(z\sim 0) = \alpha (1+z)^\beta$  and the $\pm$ $2\sigma$ range for exponent $\beta$ (if $\alpha$=1). The effective radius (proxy for size) depends strongly on redshift/lookback time.}
\label{fig7}
\end{figure*}

Figure \ref{fig7} traces the scaling of the effective radius (R$_e$) with stellar mass (left) and the evolution of R$_e(z)$/R$_e(z \simeq 0)$ as a function of redshift or, equivalently, lookback time. The SDSS and SHELS F2 stellar mass limited samples are from Figure \ref{fig5}. Again the color-coded high redshift points correspond to the samples defined in Figure \ref{fig5}. Here again we use the SDSS to define the zero redshift relation between R$_e$ and stellar mass.

There is a substantial shift toward smaller R$_e$ at fixed stellar mass between the cleanly defined SDSS (red curve) and SHELS F2 (black curve) samples. The slopes are consistent with well-established $R_e-M_\ast$ relations (e.g., \citealp{Newman12, Lange16, Damjanov19}). The slightly shallower slope of the $z\sim 0$ relation (consistent with \citealp{Shen03}) is the result of the absence of small galaxies with $M_\ast\sim2\times10^{10}\, M_{\odot}$. The R$_e$ values for the high redshift points are shifted toward even smaller radius at fixed stellar mass. Although the normalization of the relations changes substantially with the characteristic redshift of the sample, the slope appears to be insensitive to the epoch. All of these conclusions agree with well-tested previous results \citep{Newman12, Delaye14, vdW14, Lange16, Damjanov19}.

The right panel of Figure \ref{fig7}  summarizes the obvious changes in the normalization of the scaling relation that are clear in the left panel. The solid black line shows a fit to the median evolution (points with error bars) of the form  $\rm{R}_e(z)/\rm{R}_e(z\sim 0)=\alpha (1+z)^\beta$ along with the 2$\sigma$ error range for $\alpha = 1$. The evolution in R$_e$ is substantial and it is tightly constrained by the data. The resultant fit, $\beta = -1.38 \pm 0.04$ in excellent agreement with the 3DHST-CANDELS analysis of \citet{vdW14}.

As in the case of the evolution of the stellar velocity dispersion with redshift, we caution that the high redshift samples are not complete in stellar mass. The size evolution appears so robust that it may be insensitive to the selection issues. In fact, if we remove the SHELS F2 sample, the $R_e$ evolution remains unchanged. Nonetheless, here also a wider range of well-controlled samples would provide further insights including the ability to test the evolution as a function of stellar mass.

\section {The Dynamical Mass}
\label{dynmass}

Many previous studies compare dynamical and stellar masses for the quiescent population \citep{vanDokkum2009,Toft2012,Bezanson2013,vdS13,Beifiori2014,Belli2014,Belli17,Tortora2018,Stockmann2020,Mendel2020,Esdaile2021}. Generally these studies concentrate on the SDSS and high redshift samples. Here we examine the insights provided by the intermediate redshift SHELS F2 sample.

We first compare measures of the dynamical mass based on SDSS and SHELS F2 data (Section \ref{SFdmass}). We then compare dynamical and stellar masses for the SDSS, SHELS F2, and high redshift samples (Section \ref{dsmass}). In Section \ref{dsmass} we also explore the evolution of the dynamical mass - stellar mass relation.

\subsection{SDSS and F2 Measures of the Dynamical Mass}
\label{SFdmass}
The dynamical mass combines the velocity dispersion, $\sigma_e$ and the radius, $R_e$ to construct the mass proxy
\begin{eqnarray}
{\rm M}_{dyn} = \frac{K(n) \sigma_e^2 R_e}{G},
\label{mdyn}
\end{eqnarray}
where, in general, $K(n)$ is a function of the Sersic index, $n$ (e.g., \citealp{Cappellari06}). On the basis of anisotropic Jeans modeling \citep {Cappellari06} derive the following expression to account for the non-homology of the quiescent population:
\begin{eqnarray}
K_{JAM}(n) = 8.87 -0.831n + 0.0241n^2.
\label{Kn}
\end{eqnarray}

\begin{figure*}
\centering
\includegraphics[scale=0.3]{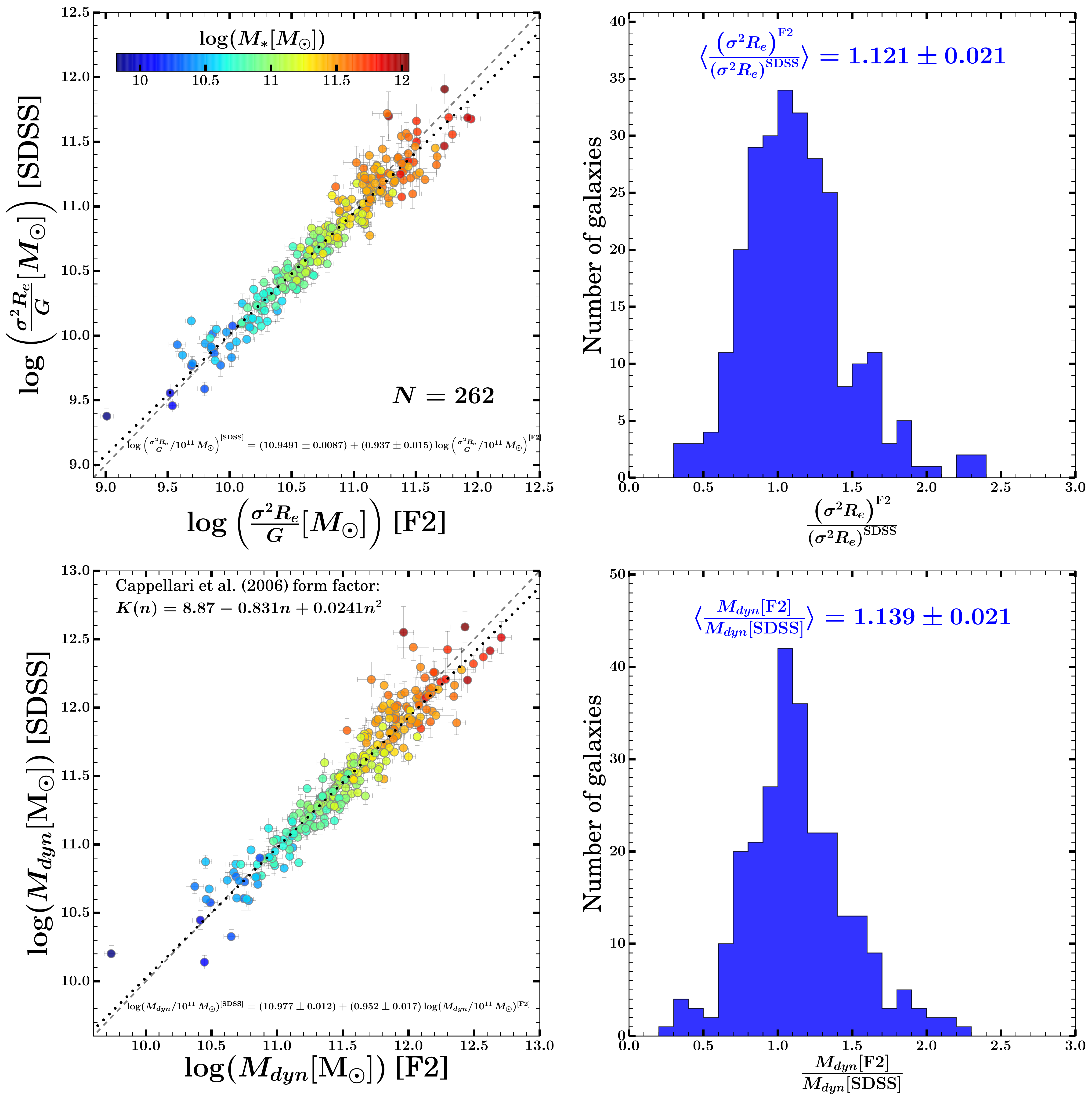}
\caption{{\it Top left:} Relation between SDSS and F2 estimates of eqn. \ref{mdyn} with $K(n) =1$ for 263 overlapping galaxies. Symbols are color-coded by galaxy stellar mass. {\it Top right:} Distribution of the ratio ({\it Top right}) between the SDSS and F2 estimates. The legend gives the average ratio and its error. {\it Bottom left:} Dynamical masses (i.e., eqn. \ref{mdyn} including the impact of the Sersic index, i.e. $K(n) \ne 1$). {\it Bottom right:} Ratio of SDSS and F2 dynamical masses; the legend gives the mean ratio and its error.}
\label{fig8}

\end{figure*}

To assess the impact of measurement error on ${\rm M}_{dyn}$ we compare SDSS and Hectospec/HSC measurements for 263 galaxies that have both SDSS and F2 data in Figure \ref{fig8}. The upper left panel of Figure assumes $K(n) = 1$ in eqn. \ref{mdyn} for the two surveys. In this comparison, the value of the Sersic index is obviously irrelevant. The slope of the obviously tight relation is slightly shallower than 1 ($0.93\pm0.02$), driven by objects with high stellar mass (color-coded according to the figure legend).

The upper right panel shows the distribution of the ratio between the SDSS and F2 estimates. The mean slightly exceeds unity because of a small number of outliers where the F2 estimate significantly exceeds the SDSS measure. The small offset from unity is irrelevant for the following analysis.

The lower two panels provide a test of the sensitivity of our overall analysis to the determination of $K(n)$. The lower left panel shows ${\rm M}_{dyn}$ for SDSS versus F2. For the SDSS we take the value of $K(n)$ that corresponds to the Sersic index in the NYU VAC. For the F2 data, the Sersic index  (Table \ref{Table3}) may be affected by systematic errors that are a function of galaxy surface brightness and the concentration of its light profile \citep[e.g.,][]{haussler_gems_2007}. The right hand histogram of the ratio between the SDSS and F2 dynamical masses is again slightly offset from a mean of 1, but the offset is within 1$\sigma$ of the offset with $K(n)$ = 1 (upper right panel). 

The slope for SDSS relative to SHELS F2 dynamical masses that takes the Sersic index into account is 0.945$\pm$0.018, close to unity. Again the high stellar mass end drives the  small departure from unity. We include $K(n)$ from \citet{Cappellari06} in all of the following analysis unless otherwise indicated. This approach takes the non-homology of the quiescent population into account \citep[see, e.g.,][] {Taylor10,Zahid17,Belli17,Mendel2020}.

\subsection {Comparing Dynamical and Stellar Masses}
\label{dsmass}

For a comparison of dynamical and stellar masses we use the $K_{JAM}(n)$ \citep{Cappellari06} based on anisotropic Jeans modeling (Eqns.~\ref{mdyn}~and~\ref{Kn}). In principle, this approach should yield a dynamical mass equal to the total luminous mass.

\begin{figure*}[h!]
    \centering
    \includegraphics[width=0.9\textwidth]{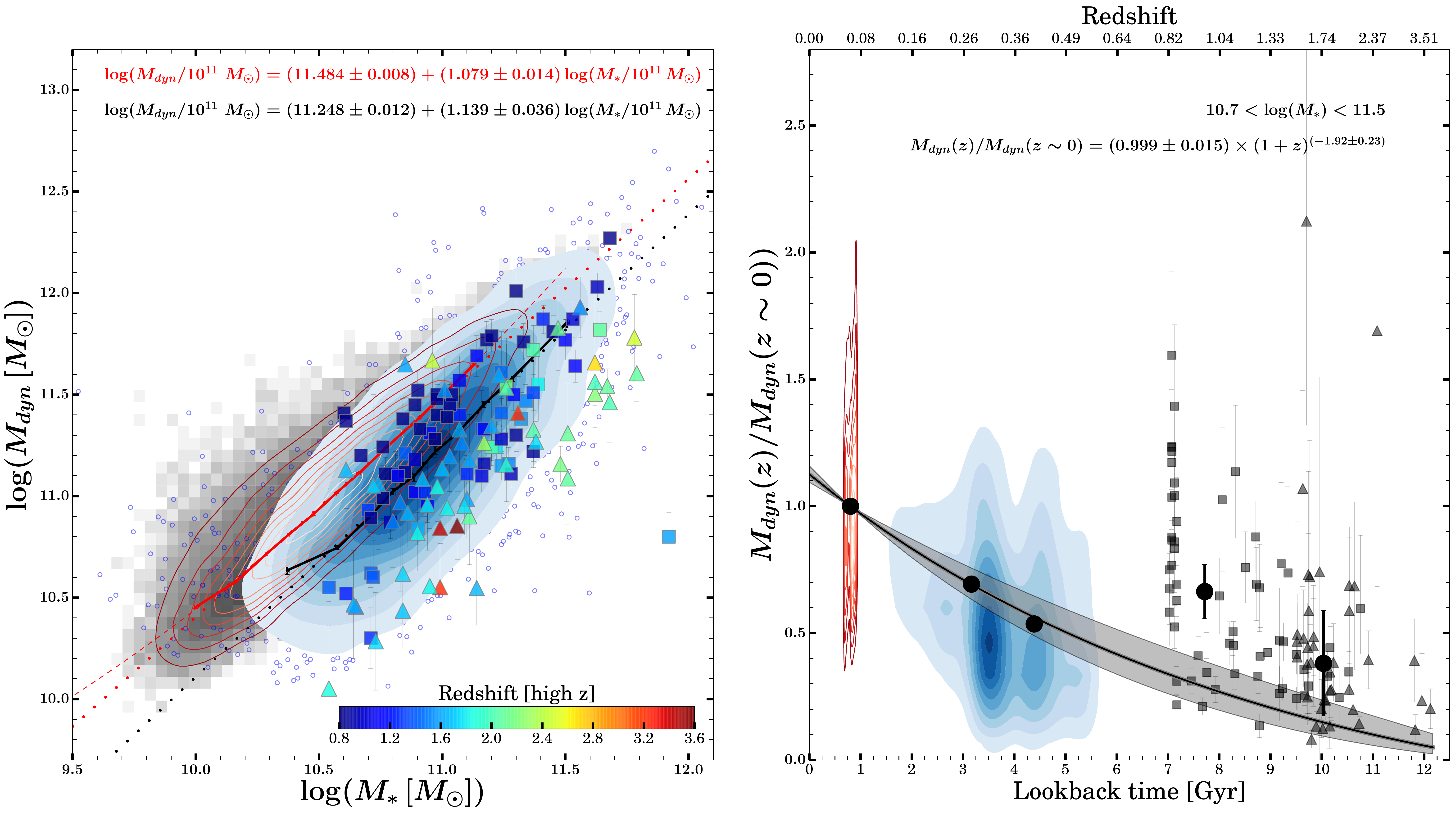}
    \caption{Dynamical vs. stellar mass for four samples: $0.05<z<0.07$ (SDSS, grey 2D histogram and red  contours of constant density), $0.1<z<0.6$ (SHELS F2, blue points and  contours), and the high redshift samples from Figure \ref{fig5}. The red (black) solid line connects median dynamical mass measurements for SDSS (SHELS F2) in equally populated mass bins. The red dashed line denotes the interpolation function (shown in extrapolation intervals for clarity) used to calculate $M_{dyn}(z)/M_{dyn}(z\sim 0)$ for higher redshifts. Dotted lines and the legend show the best linear fits to the median dynamical - stellar mass relation for SDSS (red) and  SHELS F2 (black). The offset is consistent with an increase in dynamical mass by $\sim80\%$ for a galaxy with stellar mass of $M_\ast=10^{11}\, M_\odot$. (Right) Evolution in dynamical mass at fixed stellar mass over 12~Gyrs of cosmic time (i.e., from $z\sim0$ to $z\sim3.5$ on the upper axis). Red (blue) contours show the SDSS (SHELS F2) samples. Circles with error bars represent median ratios in different redshift bins. Squares and triangles indicate the dynamical mass ratios for individual  high-$z$ galaxies as in the left-hand panel. The black solid line and shaded region show the best-fit $\ M_{dyn}(z)/M_{dyn}(z\sim 0)=\alpha \times (1+z)^\beta$ and the $\pm$ $2\sigma$ range for  $\beta$ (if $\alpha$=1).}
\label{fig9}
\end{figure*}

Figure \ref{fig9} shows the result for the samples we consider. The fits to the mass-limited SDSS (red dotted line) and SHELS F2 (black dotted line) have consistent slopes that only slightly exceed unity. The small departures of the slope from unity may result from systematic effects including but not limited to metallicity variations that affect the stellar masses and small systematic errors in the velocity dispersion measurements as a function of redshift that translate to systematic error as a function of stellar mass in a magnitude limited sample (see Figure \ref{fig3}).

The most striking aspect of the relations for all of the samples is the continual shift toward a larger ratio between dynamical mass and stellar mass as the median redshift of the sample decreases. The increase in the dynamical-to-stellar mass ratio between the median redshift of SHELS F2 ($z\sim0.35$) and SDSS ($z\sim0$) samples is $\sim60-70\%$ at constant galaxy stellar mass of $10^{11}\, M_\odot$. This change is very similar to the change in the sizes in Figure \ref{fig7}. 

\citet{Zahid17} derive the offset between dynamical and stellar mass for a different subset of the SDSS. Their offset between dynamical and stellar mass is $\sim 0.3$ dex rather than the $\sim 0.5$ dex we obtain because they use the \citet{Bertin96} relation to compute the dynamical mass. 

An offset between the dynamical and stellar mass can arise from several sources in addition to a difference in the coefficient $K(n)$ (eqn. \ref{Kn}). First, the stellar mass may be underestimated and the underestimate could be a function of redshift. In this case, the data could indicate a heavier IMF \citep[see e.g., Figure~2 in][]{Tortora2012}. Second, the difference could reflect the  contribution of dark matter to the mass within $R_e$. Obviously, without additional constraints (e.g from strong or weak lensing mass estimates), there is a degeneracy between the systematic uncertainty in the stellar mass and the dark matter mass fraction. We discuss these issues further in Sections \ref{interpretation} and \ref{lensing}.

Because we find essentially no evolution in the central stellar velocity dispersion with redshift we expect changes in the dynamical mass with redshift to be driven almost exclusively by the evolution in size. The right-hand panel of Figure \ref{fig9} shows the evolution including the high redshift samples from Figure \ref{fig5}. The black solid line and shaded region correspond to the best-fit $\rm{M}_{dyn}(z)/\rm{M}_{dyn}(z\sim 0)=\alpha (1+z)^\beta$ and the $\pm$ $2\sigma$ range for exponent $\beta$ (if $\alpha$=1). The value of $\beta = -1.92\pm 0.23$ is consistent with result for the size evolution (Figure \ref{fig7}) but is more poorly constrained. Here again our results are driven mainly by the combined SDSS and F2 mass limited samples.

Because they find that the velocity dispersion decreases as the universe ages, \citet{vdS13} find a negligible variation of $M_{dyn}/M_*$ on cosmological epoch. In contrast, using a sample of 15 $z\sim2$ galaxies with $M_\ast>10^{11}\, M_\sun$, \citet{Stockmann2020} show that in this redshift regime  $M_{dyn}/M_\ast$ ratio increases with decreasing redshift. The ratio doubles between $z\sim2$ and $z\sim0$, consistent with our results.  \citet{Esdaile2021} use a different set of 4 massive galaxies with $z \gtrsim 3$ to conclude that the ratio of $M_{dyn}/M_*$ increases to 0.33$ \pm 0.08$ dex at the current epoch, again in essential agreement with our analysis and with \citet{Zahid17}. In a recent study, using a compilation of $\sim60$~quiescent galaxies at $1.4<z<2.1$, \citet{Mendel2020} find a $+0.20\pm0.05$~dex difference between the dynamical-to-stellar mass ratios at $z\sim0$ and at high redshift. This result is in excellent agreement with the $0.24\pm0.02$~dex increase in $M_{dyn}/M_\ast$ between our SHELS F2 galaxies and the SDSS sample. 
 
Differences in the observed rate of evolution in $M_{dyn}/M_\ast$ are at least partially driven by the differences in the stellar mass range and the lack of a high redshift mass limited samples. Variations among these results suggest that additional constraints are required to unravel the complex interplay of the physical and observational issues including but not limited to quiescent galaxy selection and definition of radii \citep[see, e.g.,][]{Belli17} that underlie these results. We discuss prospects for the future in Sections~\ref{lensing}~and~\ref{conclusion}.

\section{Discussion}
\label{discussion}

The contrast between the robust size evolution and the more subtle evolution of the velocity dispersion with redshift presents a puzzle. We begin with a discussion of some of the issues including the importance of dense, complete redshift surveys covering a range of epochs in Section \ref{smevolution}. We then turn to the interpretation of the evolution of the relationship between the dynamical and stellar mass in terms of the impact of minor mergers on the dark matter content within $R_e$ (Section \ref{interpretation}). Finally we discuss the potential of gravitational lensing, both weak and strong, for more robust determination of the evolution of velocity dispersion and the dynamical-to-stellar mass ratio (Section \ref{lensing}).

\subsection{Contrast Between Size and Velocity Dispersion Evolution}
\label{smevolution}

The meaning of the word ``evolution" applied to the quiescent galaxy population can differ among various studies. In the analysis above we use the word to mean the difference between mass limited samples of quiescent objects (when available) at different redshifts. This approach does not treat progenitor bias explicitly. A few studies attempt to take progenitor bias into account explicitly, but the task is both complex and subtle.

Regardless of the details of the approach to evaluating galaxy size growth, all related studies conclude that there is significant size evolution of the quiescent population. Representing the size growth as $(1 + z)^\beta$, the value of $\beta$ for $10\lesssim\log(M_\ast/M_\sun)\lesssim 11.5$ quiescent systems spans the range   $-1\lesssim\beta\lesssim-1.5$, depending on the details of the analysis \citep{Williams2010,Damjanov2011,Ryan2012,Cassata2013,Huertas-Company2013,vdW14,Faisst2017,Mosleh2020,Yang2021}. The coefficient $\beta$ is generally less negative if the analysis includes a treatment for progenitor bias.

\citet{Zahid17} explain the impact of corrections for progenitor bias. For a given stellar mass, older objects have smaller radii and correspondingly larger velocity dispersions at a fixed epoch. They also tend to have a S\'eric index $n>3$ \citep[Figure~6 of][]{Zahid17}. The relations between the observed quantities and stellar mass are approximately parallel for different stellar population ages. Removing younger objects at more recent epochs decreases the typical size of objects in the recent epoch population. The evolution with epoch then appears less steep.

The velocity dispersion evolution is a more subtle and results in the literature range from essentially no evolution \citep[e.g.,][]{Cenarro2009, Stockmann2020} to a small but significant decrease in the dispersion as the universe ages \citep[$\sigma_e\propto(1+z)^\beta$ with $\beta\sim0.4$,][]{vdS13,Cannarozzo20}. The analysis of evolution in $\sigma_e$ for quiescent systems at $z<0.9$ in the ESO Distant Cluster Survey \citep[EDisCS,][]{Saglia_2010} results in a range of values ($0.2\lesssim\beta\lesssim0.4$) depending on, for example, whether $\beta$ is calculated at fixed dynamical mass \citep[as in][]{vdS13} or stellar mass. Regardless of the approach, \citet{Saglia_2010} show that all $\beta$ values are consistent with no evolution (within $\sim2\sigma$) in the velocity dispersion of the quiescent galaxy population. The similar constraint we obtain is driven by the combined SDSS and SHELS F2 samples; both are mass limited.

We emphasize that a mass limited sample is not velocity dispersion limited \citep[see e.g.,][]{Zahid16}. However, we are confined to mass limited samples because we cannot extract sizeable samples to the same dispersion limit for both the SDSS and SHELS F2. 

Possible systematics as a function of redshift are another observational issue. At higher redshift subtle observational issues may artificially increase the velocity dispersion (see e.g., Figure \ref{fig3}). Effects may arise from the signal-to-noise of the spectrum and the spectral resolution of the instrument.

Accounting for progenitor bias may be an issue for interpreting the evolution of the central stellar velocity dispersion as well as for the size \citep[e.g.,][]{Beifiori2014,Belli2014,Belli17,Mendel2020}. Again the work of \citet{Zahid17} provides a guide. Removal of younger and generally lower dispersion objects from the samples at more recent epochs boosts the velocity dispersion relative to measurements at earlier epochs. In the case of the SDSS and SHELS F2 samples, taking progenitor bias into account would make the velocity dispersion increase slightly toward lower redshift.

We conclude that a clean determination of the magnitude and direction of velocity dispersion evolution requires very large, complete datasets exceeding the SHELS F2 redshift range. We also conclude that the evolution in dispersion is certainly modest compared with the evolution in size. Size evolution is the primary driver of the evolution in dynamical mass that we find.

\subsection{Interpreting the Evolution of the Dynamical to Stellar Mass Relation}
\label{interpretation}

In the literature, definitions of the coefficient of the virial equation ($K(n)$ in eqn. \ref{mdyn}) vary with the particular application. \citet{Frigo17} discuss the relevant issues clearly in Section 2 of their paper. Following their notation, we take

\begin{eqnarray}
K_* = {G M_{L, r < r_{cut}}\over{R_e \sigma_e^2}},
\end{eqnarray}
where $r_{cut}$ is the cutoff radius for computation of the luminous mass, $M_L$. When there is no dark matter in the central surface brightness region of the galaxy $K_*$ = $K_{JAM}$. 

\citet{Frigo17} provide a set of expressions for $K_*$ as a function of the dark matter fraction, $f_{DM}$, within the half-light radius. They derive these expressions from models for two independent merger sequences where they follow the evolution of the stellar and dark matter components of an evolving quiescent galaxy as the progenitor undergoes successive mergers over its history. By construction, the dark matter halos of  galaxies in their simulations contain ten times the stellar mass. \citet{Frigo17} describe the details of their simulation in Section 3 of their paper.

\begin{figure*}[h!]
    \centering
    \includegraphics[width=0.9\textwidth]{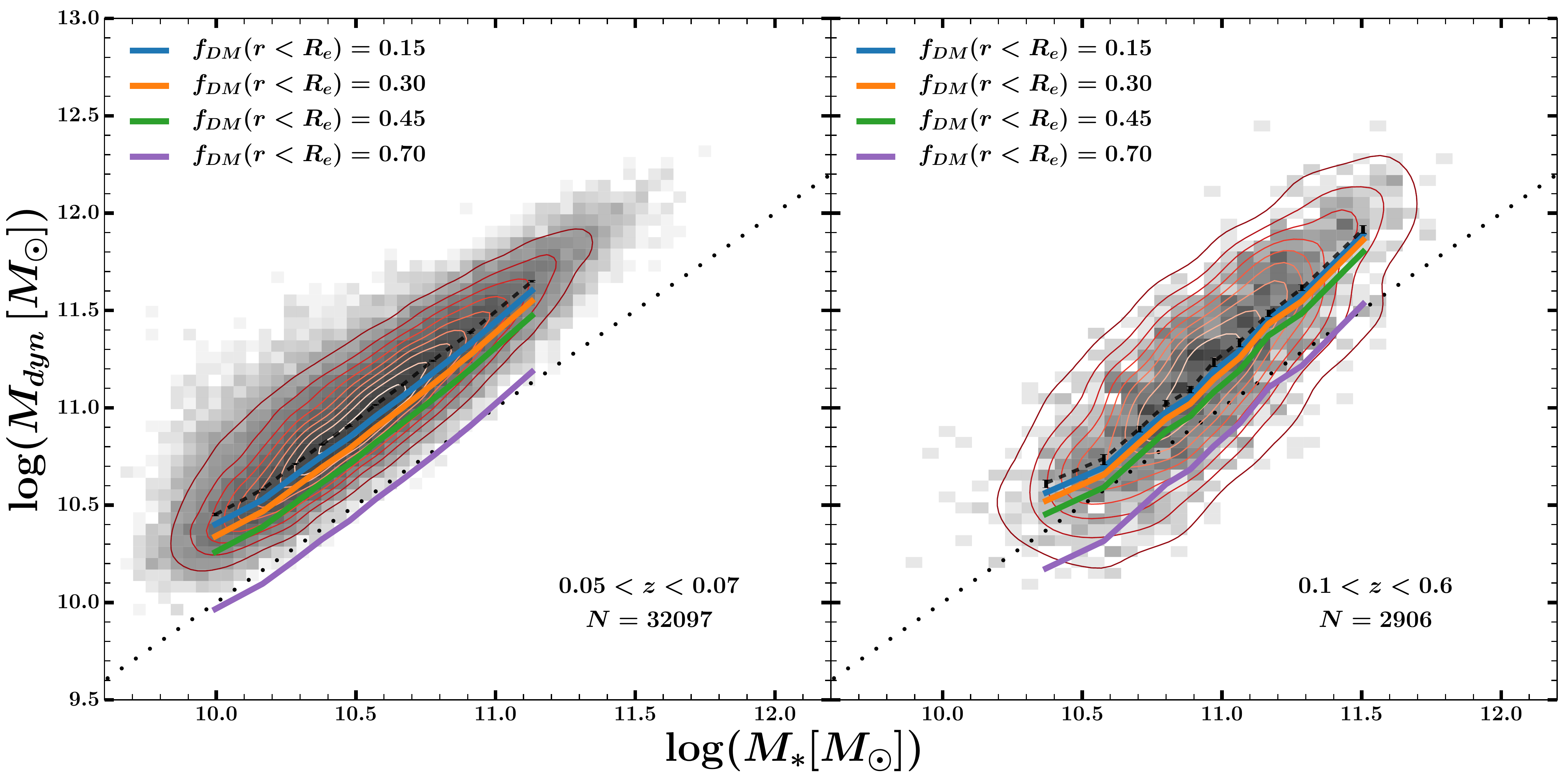}
    \caption{Dynamical vs. stellar mass for samples at $z\sim0$ (SDSS, left panel) and $0.1<z<0.5$ (SHELS F2, right panel). The grey 2D histograms and red contours show the distribution of galaxies from the two samples. Dashed black lines  connect median dynamical mass measurements ($M_{dyn}=K_{dyn}\frac{\sigma^2 R_e}{G}$) in stellar mass bins with equal number of galaxies. Colored solid lines correspond to the median total {\it stellar/luminous} mass  $M_L=K_\ast\frac{\sigma^2 R_e}{G}$ in the same stellar mass bins where $K_\ast$ depends on the fraction of dark matter within $1\, R_e$ as indicated in the legend \citep[from][Table~4]{Frigo17}. The shift of 2D distribution {\it and all} median relations with respect to 1-to-1 relation (black dotted line) from $z\sim0.35$ to $z\sim0$ suggests that the increase in dark matter fraction within $1\, R_e$  may  account  for the observed increase in $M_{dyn}/M_\ast$ ratio with decreasing redshift.}
    \label{fig10}

\end{figure*}

In Figure \ref{fig10} we use the \citet{Frigo17} models as a route toward interpreting the offset between the dynamical and stellar mass and its dependence on cosmological epoch. The two panels of Figure \ref{fig10} show the relation between dynamical and stellar mass for the SDSS (left) and SHELS F2 (right) sample with a set of superimposed curves that represent different dark matter fractions, $f_{DM}$. For each of the $f_{DM}$ in the legend we apply the formula for $K_*$ from Table 4 of \citet{Frigo17} to derive the relevant stellar/luminous mass. For $f_{DM} = 0$ (black curves) we recover the relation based on $K_{JAM}$ as expected. As $f_{DM}$ increases, the luminous mass we derive moves toward equality with the median measurements from the data.

An interesting and perhaps revealing contrast between the two different epoch samples in Figure \ref{fig10} is that at $f_{DM} = 0.7$, the derived curve lies below the data for SHELS F2, an unphysical result potentially ruled out by the data. In contrast the relation for $f_{DM} =0.7$ lies on the line of equality between derived luminous mass and measured stellar mass for the SDSS. The contrast between the two panels suggests that, as indicated by some previous investigations \citep[e.g.,][]{Beifiori2014,Tortora2018,Mendel2020}, evolution in the offset between the dynamical and stellar mass results from an increase in the dark matter fraction within $R_e$.

This interpretation remains subject to the ambiguities introduced by systematics in the computation of the stellar mass and to the degeneracy between those uncertainties and the determination of the dark matter fraction. Furthermore, the models of \citet{Frigo17} contain many albeit reasonable assumptions that require more detailed tests against the data.

As many previous investigations \citep[e.g.,][]{Hopkins2009,Hilz2012,Hilz2013,Remus2017} have emphasized, the increase in the dark matter fraction with decreasing redshift is a natural consequence of the impact of minor mergers. These mergers deposit material in the outer regions of the galaxy. They grow $R_e$ but have little impact on $\sigma_e$. As $R_e$ grows, the fraction of dark matter halo included within it increases \citep{Lovell2018}. Thus the dark matter contribution to the dynamical mass increases with cosmic time.

In a series of idealized simulations of collisionless mergers of spheroidal galaxies with mass ratios that range from 1:1 (equal-mass or major mergers) to 1:10 (minor mergers or accretions), \citet{Hilz2013} find that in minor mergers of galaxies with dark matter halos, stars in satellite galaxies are efficiently stripped at large radii leading to inside-out growth of the central (more massive) galaxy. In contrast, major mergers produce only moderate size growth and a negligible increase in the dark matter fraction within the effective radius $R_e$.

For the same increase in mass as in major mergers, minor mergers double the fraction of dark matter within $R_e$ \citep{Hilz2013}. In this case the dark matter structure of the central galaxy remains almost unchanged. However, the effective radius of its luminous matter distribution increases significantly, encompassing the regions that have been dark matter-dominated since the start of the simulation. 

\citet{Remus2017} use both hydrodynamic cosmological box simulations and zoom-in re-simulations of selected dark matter halos (at higher resolution and with added baryonic physics) to reach a similar conclusion. All of the  simulations support the two-phase evolution scenario for spheroidal galaxies, summarized in Table~2 of \citet{Oser2010}. At high redshift, mass growth is dominated by in situ star formation that creates  compact central structures with a low dark matter fraction. At lower redshifts, dry mergers start to dominate galaxy mass growth; growth in size is enhanced because mass is added to the outskirts (minor mergers dominate over rare major mergers). Thus the fraction of dark matter in the central galaxy increases because minor mergers drastically change the ruler for measuring distances within the galaxy (i.e., the effective radius).

\subsection{The Importance of Lensing}
\label{lensing}

Both strong and weak lensing measurements for large sample of objects spanning a significant range in stellar mass and redshift would be an important step toward elucidating the evolution of the central velocity dispersion and the ratio $M_{dyn}/M_*$ with redshift for the quiescent population. The obvious strength of lensing is that it measures the total mass. Thus these measurements could contribute to resolving the degeneracy between the dark matter fraction and systematics in stellar masses.

The Sloan Lens ACS Survey (SLACS) by \citet{Gavazzi07} provides a guide to the power of combined strong  and weak lensing observations for constraining the matter distribution of quiescent galaxies on the scale from 1 to 100 $R_e$. They show that for 22 massive galaxies at median redshift of $\sim 0.2$, dark matter begins to dominate the matter distribution on scales $\gtrsim R_e$. Within $R_e$ their data place an upper limit of 27$\pm$4\%. It is interesting that this limit is consistent with the indications in the right hand panel of Figure \ref{fig10} that covers a redshift range similar to the sample of \citet{Gavazzi07}. More extensive sample of strong lensing quiescent objects would obviously provide enhanced insights.

Large weak lensing studied provides an additional route toward measuring the dark matter halo velocity dispersion that corresponds very well with the central stellar velocity dispersion \citep{van_uitert_stellar_2013}. \citet{Utsumi20} use the Subaru data discussed in Section \ref{sizes} to analyze the weak lensing signal for the SHELS F2. Like \citet{Gavazzi07} and \citet{Bolton08}, \citet{Utsumi20} find that an isothermal sphere is a suitable representation of the matter density profiles for quiescent galaxies on scales $\gtrsim 100$ kpc. Furthermore, the lensing derived velocity dispersion is essentially identical to the central stellar velocity dispersion. \citet{Utsumi20} also recover a slope of $\sim 0.3$ (see Figure \ref{fig6}) for the scaling relation between the lensing derived velocity dispersion and stellar mass.

The SHELS sample analyzed here and by \citet{Utsumi20} is not large enough to use weak lensing to explore the dependence of the halo velocity dispersion of objects of fixed stellar mass at different redshifts. A much larger, deeper sample would provide a direct weak lensing test of the evolution of the central velocity dispersion with redshift perhaps resolving some of the puzzling differences discussed in Section \ref{scaleevolution}. 

\section{Conclusions}\label{conclusion}

The scaling relations and the relation between dynamical and stellar mass characterize quiescent population at a fixed redshift. Changes in the scaling relations and the dynamical-to-stellar mass ratio as a function of redshift discriminate between passive and merger-driven evolution of quiescent population. Galaxy samples that are complete in stellar mass are critical foundation for evaluation of both scaling relations and their evolution.   

We use a spectroscopic survey of the 4~deg$^2$ field (SHELS F2) and associated high-resolution HSC $i-$band images to select a mass limited sample of quiescent (D$_n4000>1.6$) galaxies with $0.1<z<0.6$. The data include stellar mass, velocity dispersion, size, and S\'{e}rsic index measurements. We provide new velocity dispersion measurements for 2985 SHELS F2 quiescent galaxies based on their MMT/Hectopsec spectra (Table~\ref{Table3}). By combining SHELS F2 galaxies with the mass limited subset of equivalently selected quiescent SDSS (DR7) galaxies at $0.05<z<0.07$, we trace the evolution in galaxy velocity dispersion ($\sigma_e$), size ($R_e$), and dynamical mass ($M_{dyn}$) at fixed stellar mass over $\sim5$~Gyrs of cosmic time. 

The large redshift range $0.1 < z < 0.6$ is critical for demonstrating that the scaling relation between velocity dispersion and stellar mass for quiescent galaxies does not evolve between $z\sim0$ and intermediate redshift (the left panel of Figure~\ref{fig6}). Over the larger redshift range the large intermediate redshift dataset constrains the evolution in velocity dispersion at fixed galaxy stellar mass for $10.7<\log(M_\ast/M_\sun)<11.5$ galaxies to be negligible. Some higher redshift data combined only with $z\sim0$ SDSS data and sparser intermediate redshift datasets imply a modest increase in the velocity dispersion at earlier epochs. A large intermediate redshift sample at $z\sim0.8-1$  would contribute substantially to resolving this apparent tension.  

The normalization of the quiescent galaxy size-stellar mass relation (left panel of Figure~\ref{fig7}) changes significantly between $z\sim0$ and $z\sim0.35$ (the median redshift of the SHELS F2 sample), confirming the results from a suite of previous studies. In contrast with the evolution in velocity dispersion, the evolution in galaxy size ($R_{e}$) is so substantial that the high redshift data are consistent with the extrapolation based on the lower redshift samples. Typically, $M_\ast\sim10^{11}\, M_\sun$ quiescent galaxies at $z\sim0.5$ (i.e., 5~Gyrs ago) were $\sim40\%$ smaller than they are today.

In combination with the shapes of galaxy light profiles (quantified by S\'{e}rsic index $n$), the observed trends in velocity dispersion and galaxy size translate into constraints on the change in the dynamical-to-stellar mass ratio relation with cosmic time (Figure~\ref{fig9}). The trend in galaxy size is the main contributor to the evolution in the dynamical-to-stellar mass ratio, which increases with decreasing redshift. At $z\sim0.5$, quiescent galaxies with stellar mass of $M_\ast\sim10^{11}\, M_\sun$ have dynamical masses ($M_{dyn}\sim1.5\times10^{11}\, M_{\odot}$) that is a factor of two lower than $M_{dyn}$ for galaxies with the same stellar mass at $z\sim0$. Our analysis of this issue agrees with the most recent high redshift results \citep{Stockmann2020,Mendel2020,Esdaile2021}. Comparison with galaxy evolutionary models that include a series of mergers with different mass ratios \citep{Hilz2013,Frigo17,Remus2017} suggests that the increase in the dynamical-to-stellar mass ratio with cosmic time is driven by the increase in the fraction of dark matter enclosed within growing $R_e$. 

Resolving the issues apparent in the velocity dispersion evolution is important for comprehensive understanding of the quiescent galaxy evolution. Routes to resolution include large mass limited samples at $z\sim0.8-1$ along with weak lensing observations. Weak lensing observations offer an independent measurement of the velocity dispersion as a function of stellar mass and redshift \citep{Utsumi20}. Existing imaging surveys like HSC SSP \citep{aihara_hyper_2018} combined with the Prime Focus Spectrograph \citep[PFS,][]{takada_extragalactic_2014} observations can provide these data.

\vspace{5mm}

I.D. acknowledges the support of the Canada Research Chair Program and the Natural Sciences and Engineering Research Council of Canada (NSERC, funding reference number RGPIN-2018-05425). J.S. is supported by the CfA Fellowship. Y.U. is supported by the U.S. Department of Energy under contract number DE-AC02-76-SF00515. M.J.G. acknowledges the Smithsonian Institution for support. I.D.A. gratefully acknowledges support from DOE grant DE-SC0010010 and NSF grant AST-2108287. This research has made use of NASA’s Astrophysics Data System Bibliographic Services. 

Funding for the SDSS-IV has been provided by the Alfred P. Sloan Foundation, the U.S. Department of Energy Office of Science, and the Participating Institutions. SDSS-IV acknowl edges support and resources from the Center for High Performance Computing at the University of Utah. The SDSS website is \url{www.sdss.org}. SDSS-IV is managed by the Astrophysical Research Consortium for the Participating Institutions of the SDSS Collaboration including the Brazilian Participation Group, the Carnegie Institution for Science, Carnegie Mellon University, Center for Astrophysics | Harvard and Smithsonian, the Chilean Participation Group, the French Participation Group, Instituto de Astrof\'{i}sica de Canarias, Johns Hopkins University, Kavli Institute for the Physics and Mathematics of the Universe (IPMU)/University of Tokyo, the Korean Participation Group, Lawrence Berkeley National Laboratory, Leibniz Institut f\"{u}r Astrophysik Potsdam (AIP), Max-Planck-Institut f\"{u}r Astronomie (MPIA Heidelberg), Max- Planck-Institut f\"{u}r Astrophysik (MPA Garching), Max-Planck- Institut f\"{u}r Extraterrestrische Physik (MPE), National Astronomical Observatories of China, New Mexico State University, New York University, University of Notre Dame, Observat\'{a}rio Nacional/MCTI, Ohio State University, Pennsylvania State University, Shanghai Astronomical Observatory, United Kingdom Participation Group, Universidad Nacional Aut\'{o}noma de M\'{e}xico, University of Arizona, University of Colorado Boulder, University of Oxford, University of Portsmouth, University of Utah, University of Virginia, University of Washington, University of Wisconsin, Vanderbilt University, and Yale University.

The Hyper Suprime-Cam (HSC) collaboration includes the astronomical communities of Japan and Taiwan as well as Princeton University. The HSC instrumentation and software were developed by the National Astronomical Observatory of Japan (NAOJ), the Kavli Institute for the Physics and Mathematics of the Universe (Kavli IPMU), the University of Tokyo, the High Energy Accelerator Research Organization (KEK), the Academia Sinica Institute for Astronomy and Astrophysics in Taiwan (ASIAA), and Princeton University.
Funding was contributed by the FIRST program from the Japanese Cabinet Office, the Ministry of Education, Culture, Sports, Science and Technology (MEXT), the Japan Society for the Promotion of Science (JSPS), Japan Science and Technology Agency (JST), the Toray Science Foundation, NAOJ, Kavli IPMU, KEK, ASIAA, and Princeton University. This paper makes use of software developed for the Large Synoptic Survey Telescope (LSST). We thank the LSST Project for making their code available as free software at \url{http://dm.lsst. org}. This paper is based [in part] on data collected at the Subaru Telescope and retrieved from the HSC data archive system, which is operated by Subaru Telescope and Astronomy Data Center (ADC) at National Astronomical Observatory of Japan. Data analysis was in part carried out with the cooperation of Center for Computational Astrophysics (CfCA), National Astronomical Observatory of Japan.

The authors wish to recognize and acknowledge the very significant cultural role and reverence that the summit of Maunakea has always had within the indigenous Hawai'ian community. We are most fortunate to have the opportunity to conduct observations from this sacred mountain.

\software{SciPy \citep{2020SciPy-NMeth}, NumPy \citep{2020NumPy-Array}, AstroPy \citep{collaboration_astropy_2013,astropy_collaboration_astropy_2018}, Scikit-learn \citep{pedregosa2011scikit}, Matplotlib \citep{Hunter:2007}, Seaborn \citep{Waskom2021}}


\end{document}